\pdfoutput=1
\documentclass[fleqn,usenatbib,useAMS,onecolumn]{mnras}

\usepackage{graphicx}	
\usepackage{amsmath}	
\usepackage{amssymb}	
\usepackage{multicol}        
\usepackage{bm}		
\usepackage{pdflscape}	
\usepackage{epstopdf}
\usepackage[T1]{fontenc}
\usepackage{ae,aecompl}

\newcommand{\ms}[1]{\textcolor{black}{#1}}
\newcommand{\msrev}[1]{\textcolor{black}{#1}}

\newcommand{\beq}{\begin{equation}}
\newcommand{\beqa}{\begin{eqnarray}}
\newcommand{\eeq}{\end{equation}}
\newcommand{\eeqa}{\end{eqnarray}}

\newcommand{\simgt}{\lower.5ex\hbox{$\; \buildrel > \over \sim \;$}}
\newcommand{\simlt}{\lower.5ex\hbox{$\; \buildrel < \over \sim \;$}}

\newcommand{\bx}{{\bf x}}

\newcommand{\bk}{{\bf k}}
\newcommand{\bmtheta}{{\bm{\theta}}}

\newcommand{\tdelta}{\tilde{\delta}}
\newcommand{\tkappa}{\tilde{\kappa}}

\newcommand{\dsigma}{{\Delta\Sigma}}

\newcommand{\sigmacr}{\Sigma_{\rm cr}}
\newcommand{\sigmacri}{\Sigma_{\rm cr}^{-1}}

\def\avrg#1{\left\langle #1 \right\rangle}

\usepackage[T1]{fontenc}
\usepackage{ae,aecompl}

\usepackage{newtxtext,newtxmath}

\title[$\Delta\Sigma$ vs. $\gamma_+$ for stacked lensing]{
Stacked lensing estimators and their covariance matrices: 
Excess surface mass density vs. Lensing shear
}

\author[M. Shirasaki \& M. Takada]
{
Masato Shirasaki$^{1}$\thanks{Contact e-mail: \href{mailto:masato.shirasaki@nao.ac.jp}{masato.shirasaki@nao.ac.jp}}
and
Masahiro Takada$^{2}$
\\
$^{1}$Division of Theoretical Astronomy, National Astronomical Observatory of Japan, 
Mitaka, Tokyo 181-8588, Japan \\
$^{2}$Kavli Institute for the Physics and Mathematics of the Universe
(WPI), The University of Tokyo Institutes for Advanced Study (UTIAS), \\
\,
The University of Tokyo, Chiba 277-8583, Japan
}

\date{}

\pubyear{2018}

\begin{document}
\label{firstpage}
\pagerange{\pageref{firstpage}--\pageref{lastpage}}
\maketitle

\begin{abstract}
Stacked lensing is a powerful means of measuring 
the average mass distribution around large-scale structure tracers.
There are two stacked lensing estimators used in the literature, denoted as $\Delta\Sigma$ and $\gamma_+$, 
which are related as $\Delta\Sigma=\Sigma_{\rm cr}\gamma_+$, where 
$\sigmacr(z_l,z_s)$ is the critical surface mass density for each lens-source pair 
($z_l$ and $z_s$ are lens and source redshifts, respectively). 
In this paper we derive a formula for the covariance matrix of $\Delta\Sigma$-estimator 
focusing on ``weight'' function to improve the signal-to-noise ($S/N$).
We assume that the lensing fields and the distribution of lensing objects obey the Gaussian statistics. With this formula, we show that, if background galaxy shapes are weighted by an amount of $\sigmacr^{-2}(z_l,z_s)$, the $\Delta\Sigma$-estimator maximizes the $S/N$ in the shot noise limited regime.
We also show that the $\Delta\Sigma$-estimator with the weight 
$\sigmacr^{-2}$ gives a greater $(S/N)^2$ than that of the $\gamma_+$-estimator by about 5--25\% for lensing objects at redshifts comparable with or higher than the median of source galaxy redshifts for hypothetical Subaru HSC and DES surveys. However, for low-redshift lenses such as $z_l\simlt 0.3$, 
the $\gamma_+$-estimator has higher $(S/N)^2$ than $\Delta\Sigma$. 
We also discuss that the $(S/N)^2$ for $\Delta\Sigma$ at large separations 
in the sample variance limited regime can be boosted, by up to a factor of 1.5, if one adopts a weight 
of $\sigmacr^{-\alpha}$ with $\alpha>2$.
Our formula allows one to explore how the combination of the different estimators can approach an optimal estimator in all regimes of redshifts and separation scales.
\end{abstract}





\section{Introduction}


Modern galaxy surveys enable us to study large-scale structures (LSS) 
in the universe with various statistical methods.
Among them, the cross-correlation of LSS tracers with shapes of background galaxies,
referred as to stacked lensing or galaxy-galaxy lensing, is a unique means of measuring the average total matter distribution around the foreground objects.
Furthermore, combining the stacked lensing and auto-clustering correlation of the same foreground
tracers, one can recover the spatial relation between the foreground tracers and the surrounding matter distribution in a statistical sense, 
and then
constrain cosmology by breaking degeneracies between bias uncertainty and cosmological parameters
\citep[e.g.,][]{Seljaketal:05,2009MNRAS.394..929C,Mandelbaumetal:13,2015ApJ...806....2M,2017MNRAS.464.4045K, 2017arXiv170605004V, 2017arXiv170801530D}.
Therefore, stacked lensing measurements are expected to be one of the most powerful probes for 
ongoing and upcoming wide-area galaxy surveys for addressing fundamental physics such as the nature of dark energy and neutrino mass 
\citep[e.g.,][]{OguriTakada:11,Schaanetal:16}.
The ongoing surveys include the Dark Energy
Survey (DES)\footnote{\url{https://www.darkenergysurvey.org/}}, 
the Kilo-Degree Survey (KiDS)\footnote{\url{http://kids.strw.leidenuniv.nl}}, 
and the Subaru Hyper Suprime-Cam (HSC) survey\footnote{\url{http://hsc.mtk.nao.ac.jp/ssp/}},
whereas we will have access to larger amounts of information from next-generation surveys such as 
the Dark Energy Spectroscopic Instrument (DESI)\footnote{\url{http://desi.lbl.gov}}, 
and the Subaru Prime Focus Spectrograph (PFS) survey \citep{Takadaetal:14} in
5-years timescale, and ultimately the Large Synoptic Survey Telescope
(LSST)\footnote{\url{https://www.lsst.org}},
Euclid\footnote{\url{http://sci.esa.int/euclid/}} and WIFRST\footnote{\url{http://wfirst.gsfc.nasa.gov}} 
in the next decade.

There are two estimators of stacked lensing that have been used in the literature;
one is the stacked shear profile, denoted as $\gamma_+$ in this paper, and the other is the excess surface mass 
density profile, denoted as $\Delta \Sigma$\footnote{
In this paper, we do not consider a different estimator known as annular differential surface density (ADSD) 
profile \citep{2010PhRvD..81f3531B, 2010MNRAS.405.2078M}.
The ADSD estimator is designed so as to make the measured projected mass profile insensitive to 
nonlinear, small-scale information.
}. 
The estimator of $\gamma_+$ is obtained by averaging tangential ellipticity components of background 
galaxies with respect to the center of foreground LSS tracers over all the lens-source pairs of a given separation on the sky. 
This has been applied to actual observational data 
\citep[e.g.][]{1996ApJ...466..623B, 1998ApJ...503..531H, Fischeretal:00, 2004ApJ...606...67H,Okabeetal:10,PratetalDESgglensing:17}.
On the other hand, if the redshift information of lensing objects and source galaxies are available, 
one can estimate the average surface mass density profile of lensing objects, $\Delta \Sigma$, 
by multiplying the critical surface mass density, $\sigmacr(z_l,z_s)$, with ellipticity of background galaxy
for each of lens-source pairs with redshifts 
$z_l$ and $z_s$, respectively, in the stacked lensing measurement.
The measurement of $\Delta \Sigma$ has been performed 
\citep[e.g.][]{2001astro.ph..8013M, 2002MNRAS.335..311G, 2006MNRAS.368..715M,2007arXiv0709.1159J, 2013MNRAS.431.1439G, 2013ApJ...769L..35O, 2014MNRAS.437.2111V,Miyatakeetal:15}.
All ongoing and upcoming wide-area weak lensing surveys carry out multi-color photometric surveys, which can
be used to estimate redshifts of each galaxy based on photometric redshift method. 
In addition, wide-area spectroscopic galaxy surveys can have an overlap with the survey footprint of weak lensing survey, for example, which
is the case for the Subaru Hyper Surpime-Cam and the SDSS survey.  
Such a spectroscopic galaxy catalog can be used to 
define a secure sample of foreground lensing objects for the stacked lensing analysis. Then 
the lensing information can be combined with the auto-correlation function and redshift-space distortion effect of the spectroscopic 
galaxies in order to improve the cosmological constraints \citep{Seljaketal:05,Reyesetal:10,2013MNRAS.435.2345H,2015ApJ...806....2M,2017arXiv170605004V,2017arXiv170801530D,2018MNRAS.474.4894J}. 
Another subtle advantage for the use of redshift information of lensing objects is we can measure the 
stacked lensing profile as a function of  projected radii from lensing objects, rather than angular scales, which does not 
mix different scales of the matter distribution even after line-of-sight projection.

Then a natural question arises; are the two estimators of stacked lensing, $\Delta\Sigma$ and $\gamma_+$, equivalent 
to each other, if those are applied to the exactly same data sets (the same lens-source pairs and the same survey area)?
If this is not the case, which estimator is optimal in terms of the signal-to-noise ratio?  
Although the tangential shear is related to the excess surface mass density as
$\gamma_+ = \Delta\Sigma/\sigmacr$ on object-by-object basis, this question, after the statistical average, is not trivial. 
Hence the main purpose of this paper is to address the above questions. 
To do this, we derive the covariance matrix for the stacked lensing profile, assuming that both the lensing fields and the 
distribution of lensing objects obey the Gaussian distributions. Whilst the covariance matrix formula for $\gamma_+$ is derived in the literature 
\citep{OguriTakada:11,KrauseEifler:17}, the covariance matrix formula for $\Delta\Sigma$ has yet to be derived, except for 
some studies using the real data and/or mock catalogs \citep{Mandelbaumetal:13,Shirasakietal:17,2016arXiv161100752S}. 
When deriving the covariance matrix, we include the ``weight'' function in the stacked lensing estimator that is often 
used in actual observations in order to extract the maximum information content from a given data set, which in turn allows to obtain 
tightest constraints on model parameters including cosmological parameters. 
The weight that is often used in the literature is $w\propto 1/(\sigma_{\rm SN}^2+\sigma_e^2)$ for $\gamma_+$
\citep[e.g.,][]{BernsteinJarvis:02,Okabeetal:10}
or $w\propto \sigmacr^{-2}/(\sigma_{\rm SN}^2+\sigma_e^2)$ for $\Delta\Sigma$
\citep[e.g.,][]{2004AJ....127.2544S,Mandelbaumetal:05,Mandelbaumetal:06,Miyatakeetal:15,Murataetal:17}, motivated 
by the inverse-variance weighting in the shape noise dominated regime, where $\sigma_{\rm SN}$ is the intrinsic rms ellipticity of source galaxies 
and $\sigma_e$ is the measurement error. For $\Delta\Sigma$ the weight $\sigmacr^{-2}$ down-weights pairs of lens-source galaxies that are 
close in redshifts and therefore have a lower lensing efficiency. 
Once the covariance matrix formula is derived, we can discuss which weight can be optimal to maximize the signal-to-noise 
ratio for the stacked lensing measurement as well as can address which estimator of $\gamma_+$ or $\Delta\Sigma$
is optimal. Our results would also help to guide an optimal planning of stacked lensing measurement for a given 
galaxy survey.   
To validate the covariance matrix formula, we will use mock catalogs of lensing halos and background galaxy shapes 
in the light cone simulations \citep{Shirasakietal:17}
by comparing the analytical prediction with the simulation results. 
 
The paper is organized as follows.
Section~\ref{sec:basics} summarizes basics of the statistical property of foreground LSS tracers and weak lensing.
Section~\ref{sec:estimator} describes the estimators of $\Delta \Sigma$ and $\gamma_+$ and derives their covariance matrices
by assuming the Gaussian statistics.
Section~\ref{sec:results} presents the main results including validation of our model with numerical simulations
as well as how we can realize the improvement of the signal-to-noise ratio of $\Delta\Sigma$ compared to $\gamma_+$.
We conclude this paper in Section~\ref{sec:conclusions}.

\section{Preliminaries}
\label{sec:basics}

\subsection{Projected number density field of halos}

Let us assume that we have a sample of halos distributed over a solid angle of the survey field. Then consider a case that we 
use this sample of halos for the stacked lensing measurement by cross-correlating positions of halos with shapes of background galaxies. 
%
In this case, the angular number density of halos per unit steradian
can be written as
\beq
n^{\rm 2D}_{\rm h}(\bmtheta)=\bar{n}^{\rm 2D}_{\rm h}\left[1+\delta_{\rm h}^{\rm 2D}\!(\bmtheta)\right],
\eeq
where $\delta_{\rm h}^{\rm 2D}(\bmtheta)$ is the projected number density fluctuation field, which is dimension-less. 
$\bar{n}^{\rm 2D}_{\rm h}$ is the mean number density, 
expressed in terms of the halo mass function as
%
\beq
\bar{n}^{\rm 2D}_{\rm h}=\int_{\chi_l-\Delta\chi_l/2}^{\chi_l+\Delta\chi_l/2}\!\mathrm{d}\chi~ \chi^2 f_{\rm h}(\chi)\int\!\!\mathrm{d}M~\frac{\mathrm{d}n}{\mathrm{d}M}S(M,\chi).
\eeq
Here $\chi$ is the radial comoving distance to redshift $z$, and is given via the distance-redshift relation for an assumed cosmological model as $\chi=\chi(z)$; 
$\chi_l$ is the mean radial distance to halos in the sample, $\Delta\chi_l$ is the width of their radial distances; 
$\mathrm{d}n/\mathrm{d}M$ is the halo mass function; $S(M,\chi)$ is the selection function of halo mass. 
\ms{In this paper, we simply define $f_{\rm h}(\chi)$ so that $f_{\rm h}(\chi)=1$ if $\chi(z)$ is in the redshift range of interest, i.e.
$\chi_l-\Delta\chi/2\le \chi\le \chi_l+\Delta\chi_l/2$, and otherwise $f_{\rm h}(\chi)=0$.}
If halos in the sample is distributed in a narrow redshift range, 
the angular number density is approximated as $\bar{n}^{\rm 2D}_{\rm h}\simeq \chi_l^2\Delta\chi_l \int\!\!\mathrm{d}M~\left.\frac{\mathrm{d}n}{\mathrm{d}M}\right|_{\chi_l}S(M,\chi)$.
Note $\bar{n}^{\rm 2D}_{\rm h}$ is dimension-less and gives the angular number density, the number of halos per unit steradian. 
In the following we omit the superscript ``2D'' in the angular number density $\bar{n}^{\rm 2D}_{\rm h}$
for notational simplicity. 
The 2D field $\delta^{\rm 2D}_{\rm h}(\bmtheta)$ is 
expressed in terms of the three-dimensional number density field of halos \ms{as} 
%
\beq
\delta^{\rm 2D}_{\rm h}(\bmtheta)\equiv 
\frac{1}{\bar{n}_{\rm h}}\int_{\chi_l-\Delta\chi_l/2}^{\chi_l+\Delta\chi_l/2}\!\!\mathrm{d}\chi~ \chi^2 f_{\rm h}(\chi)
\int\!\!\mathrm{d}M~\frac{\mathrm{d}n}{\mathrm{d}M}S(M,\chi)\delta_{\rm h}(\chi\bmtheta,\chi;M).
\eeq
%
\ms{Here we introduce the three-dimensional number density 
fluctuation field of halos, $\delta_{\rm h}$, 
via $n_{\rm h}(\bx; M) = {\rm d}n/{\rm d}M\, \left[1+\delta_{\rm h}(\bx; M)\right]$,
where $n_{\rm h}(\bx; M)$ is the three-dimensional number density field 
at the position, $\bx=(\chi,\chi\bmtheta)$, for halos of mass $M$.}
Throughout this paper we employ a flat geometry universe.

For convenience of our discussion, we introduce the following two-dimensional Fourier transform of a field in the two-dimensional plane at the radial distance $\chi_l=\chi(z_l)$, perpendicular to the line-of-sight direction: 
%
\beq
\delta^{\rm 2D}_{\rm h}\!(\bx_\perp)\equiv \int\!\!\frac{\mathrm{d}^2\bk_\perp}{(2\pi)^2}~ \tdelta^{\rm 2D}_{\rm h}\!(\bk_\perp,z_l)e^{i\bk_{\perp}\cdot\chi_l\bmtheta},
\label{eq:2DFourier}
\eeq
where $\bx_\perp\equiv \chi_l\bmtheta$. 
Throughout this paper we employ the flat-sky approximation. 
The inverse Fourier transform is given as
%
\beq
\tdelta^{\rm 2D}_{\rm h}(\bk_\perp)
\simeq \frac{1}{\bar{n}_{\rm h}}\int_{\chi_l-\Delta\chi_l/2}^{\chi_l+\Delta\chi_l/2}\!\!\mathrm{d}\chi~ \chi^2f_{\rm h}(\chi)
\int\!\!\mathrm{d}M\frac{\mathrm{d}n}{\mathrm{d}M}S(M,\chi)
\int\!\!\frac{\mathrm{d}k_\perp}{2\pi}\tdelta_{\rm h}(k_\parallel,\bk_\perp)e^{ik_\parallel\chi},
\label{eq:deltah_2d}
\eeq
where $\tdelta_{\rm h}(\bk)$ is the Fourier transform of the three-dimensional number density fluctuation field of halos, and 
we have used $\chi\simeq \chi_l$ in the redshift range of halos in the sample.
Now we consider the projected auto-correlation function of halos defined by
\begin{equation}
\xi_{\rm hh}(R)\equiv \left.\avrg{\delta^{\rm 2D}_{\rm h}(\bx_\perp)\delta^{\rm 2D}_{\rm h}(\bx_\perp^\prime)}\right|_{R=|\bx_\perp-\bx_\perp^\prime|}
\end{equation}
where the average is for all the pairs that are in the projected separation, $R$. 
Using the Limber's approximation \citep{Limber:54}, the projected correlation function of halos can be computed as
%
\beq
\xi_{\rm hh}(R)\equiv  \frac{1}{(\bar{n}_{\rm h})^2} \int_{\chi_l-\Delta\chi_l/2}^{\chi_l+\Delta\chi_l/2}\!\!\mathrm{d}\chi~
\chi^4 f_{\rm h}(\chi)^2\left[\int\!\!\mathrm{d}M\frac{\mathrm{d}n}{\mathrm{d}M}b(M)S(M,\chi)\right]^2
\int\!\!\frac{k_\perp\mathrm{d}k_\perp}{2\pi}
P^L_{\rm m}(k_\perp;\chi)J_0(kR),\label{eq:whh}
\eeq
%
where $J_0(x)$ is the zeroth-order Bessel function and
we have assumed that the three-dimensional correlation function of halos with masses $M$ and $M'$ is given by
$P_{\rm hh}(k;M,M')\simeq b(M)b(M')P^{L}_{\rm m}(k)$, where $b(M)$ is the linear bias parameter for halos of mass $M$ and $P^L_{\rm m}(k)$ is the linear matter power spectrum. From the above equation, we can define
 the projected power spectrum of the halo correlation function as
%
\beq
C_{\rm hh}(k)\equiv 
\frac{1}{(\bar{n}_{\rm h})^2} \int_{\chi_l-\Delta\chi_l/2}^{\chi_l+\Delta\chi_l/2}\!\!\mathrm{d}\chi~
\chi^4 f_{\rm h}(\chi)^2\left[\int\!\!\mathrm{d}M\frac{\mathrm{d}n}{\mathrm{d}M}b(M)S(M,\chi)\right]^2
P^L_{\rm m}(k_\perp;\chi).
\eeq
The dimension of $C_{\rm hh}(k)$ is $[({\rm Mpc})^2]$. The ``observed'' power spectrum of halos is affected by the shot noise due to a finite number of halos in the sample, and is expressed as
\beq
C_{\rm hh}^{\rm obs}(k)=C_{\rm hh}(k)+\frac{\chi_l^2}{\bar{n}_{\rm h}},
\eeq
where we have assumed a narrow redshift bin of halos for consistency with the following discussion; we will often employ this approximation for clarity of our discussion, but this is not an important 
assumption for the main purpose of this paper. The second term denotes the shot noise.

\subsection{Cosmic shear power spectrum}

Consider a sample of source galaxies from which the weak lensing effects can be measured. Suppose that the redshift distribution of 
source galaxies is given by
%
\beq
\frac{\mathrm{d}n}{\mathrm{d}z_s}dz_{s}= \bar{n}_{\rm tot}p(z_s)\mathrm{d}z_s,
\eeq
where $\bar{n}_{\rm tot}$ is the mean number density of all the source galaxies per unit steradian and $p(z_{s})$ is the normalized redshift distribution, defined
so as to satisfy $\int_0^\infty\!\!\mathrm{d}z_s~ p(z_s)=1$. Note that we used the notation $\bar{n}$ to denote the ``angular'' number 
density, 
and please do not confuse it with $\mathrm{d}n/\mathrm{d}M$, which denotes the three-dimensional 
number density of halos.
In this paper, following \cite{TakadaJain:09} and \cite{OguriTakada:11}, we 
employ the following, simplified form to model the redshift distribution:
%
\beq
p(z_s)\propto z_s^2 \exp\left[-\frac{z_s}{z_0}\right],
\label{eq:model_pz}
\eeq
where $z_0$ is a parameter to model the depth of redshift distribution (the higher $z_0$ is, the higher redshift the source distribution peaks at),
and the normalization factor is determined by the normalization condition. 
With this form, the mean redshift of galaxies is given as $\avrg{z_s}=\int\!\!\mathrm{d}z_s~ p(z_s)z_s = 3z_0$. 
To model a Subaru HSC-type survey, we choose $z_0=1/3$ so that 
the mean redshift $\avrg{z_s}=1$, while $z_0=0.7/3$ for a DES-type survey as $\avrg{z_s}=0.7$.
For the number density, $\bar{n}_{\rm tot}$, that also characterizes the depth of a given survey, we 
assume $\bar{n}_{\rm tot}=20~$arcmin$^{-2}$ and $7$~arcmin$^{-2}$ for the HSC- and DES-type surveys, respectively.  

The cosmic shear effect on a source galaxy in the angular direction $\bmtheta_s$ and at redshift $z_s$ is caused by foreground structures along the path of light ray: 
%
\beq
\kappa(\bmtheta_s,z_s)\equiv \int^{z_s}_0\!\!\mathrm{d}\chi~ \sigmacri(z,z_s) \bar{\rho}_{\rm m0}\delta_{\rm m}(\chi,\chi\bmtheta_s),
\eeq
%
where we have assumed the Born approximation, which is a good approximation for statistical quantities of weak lensing effects in which 
we are interested. 
For simplicity of the following discussion, we here consider the lensing convergence field, $\kappa(\bmtheta_s,z_s)$, rather 
than the shear field, but the two fields are equivalent (see below). $\sigmacr(z,z_s)$ is the critical surface density, defined as
%
\beq
\sigmacri(z,z_s)\equiv 
\left\{
\begin{array}{ll}
4\pi G a(z)^{-1}\chi(z)\left[1-\frac{\chi(z)}{\chi(z_s)}\right], & \mbox{if $z\le z_s$} \nonumber\\
0, & \mbox{if $z>z_s$}.
\end{array}
\right.
\eeq
and $\chi$ is the comoving angular diameter distance;  
$\chi(z)$ is equivalent to the comoving radial distance for a flat-geometry universe.
For the definition of $\sigmacr$, we followed \cite{Mandelbaumetal:13} \citep[also see][]{Miyatakeetal:15,Murataetal:17} where the critical surface mass density is defined in units of the comoving coordinates, i.e. $[\sigmacr]=[M/L_{\rm comoving}^2]$. Note that this definition differs from that in \cite{OguriTakada:11}, where the critical density is defined in units of the physical lengths: $\sigmacr^{\rm phy}=\sigmacr^{\rm com}/a^2$.

For convenience of the following discussion we consider a projection of the observed cosmic shear field to the two-dimensional flat space at a 
redshift of lensing halos, say $\chi_l\equiv \chi(z_l)$. 
Similarly to Eq.~(\ref{eq:2DFourier}), we define the following Fourier transform in the two-dimensional flat space at $z_l=z(\chi_l)$: 
%
\beq
\tkappa(\bk_\perp,z_s)\equiv \int\!\!\mathrm{d}^2\bx_\perp~ \kappa(\bmtheta_s,z_s)e^{-i\bk_\perp\cdot\chi_l\bmtheta_s},
\eeq
where $\bx_\perp\equiv \chi_l\bmtheta_s$. Using the Limber's approximation, the projected power spectrum of cosmic shear is computed, e.g. following 
\cite{TakadaJain:04} \citep[also see][]{DodelsonBook}, as
\beq
C_{\kappa\kappa}(k_\perp)\equiv \int_0^\infty\!\!\mathrm{d}\chi~ W_{\rm GL}(\chi)^2 \left(\frac{\chi_l}{\chi}\right)^2
P_{\rm m}\!\left(k=\frac{\chi_l}{\chi}k_\perp; \chi\right),
\label{eq:Ckappa}
\eeq
where the factor $(\chi_l/\chi)^2$ arises from the fact that we used the flat-space, rather than angular, Fourier transform
in the plane of lensing halos (see below), and 
$W_{\rm GL}(\chi)$ is the lensing efficiency function for the sample of source galaxies, defined as
\beq
W_{\rm GL}(\chi)\equiv 
\int_{z=z(\chi)}^\infty\!\!
\mathrm{d}z_s~p(z_s) \sigmacri(z,z_s)\bar{\rho}_{\rm m0}
=
\int_{z=z(\chi)}^\infty\!\!
\mathrm{d}z_s~p(z_s) 4\pi G a^{-1}\bar{\rho}_{\rm m0}
\chi\left[1-\frac{\chi}{\chi(z_s)}\right].
\eeq
The ``observed'' power spectrum of cosmic shear is affected by the shape noise arising from the intrinsic galaxy ellipticities as well as
a finite number of source galaxies used in the sample. The observed power spectrum is given as
\begin{equation}
C^{\rm obs}_{\kappa\kappa}(k)=C_{\kappa\kappa}(k)+\chi_l^2\frac{\sigma_\epsilon^2}{\bar{n}_{\rm tot}},
\label{eq:Ckappa_obs}
\end{equation}
where $\sigma_\epsilon$ is the rms of intrinsic ellipticity per component.
The dimension of $C_{\kappa\kappa}$ is $[({\rm Mpc})^2]$.

\section{Stacked lensing profile and the Covariance matrix}
\label{sec:estimator}
\subsection{Stacked surface mass density profile}
\label{subsec:dSigma}

Now we consider the stacked lensing that is the main focus of this paper. 
By stacking shapes of background galaxies around the lensing halos over 
all the pairs each of which is separated by the same projected radius at the lens redshift, say $R$, 
one can measure the {\em average} projected
matter density profile around the lensing halos. The estimator can be written as
%
\beq
\widehat{\avrg{\Sigma}}(R)\equiv \left.\frac{1}{N_{w,{\rm pair}}(R)}\sum_{l,s}~w_{ls}\sigmacr(z_l,z_s)\kappa(\bmtheta_s,z_s)\right|_{R=\chi_l|\bmtheta_l-\bmtheta_s|},
\label{eq:est_sigma}
\eeq
where $z_l$ and $z_s$ are redshifts of lensing halo and source galaxy in each pair, respectively. For generality of our discussion, we 
introduced a weight function, denoted by $w_{ls}$, and will discuss how the expected signal-to-noise ratio for the stacked
lensing measurement varies with a different choice of the weight function. 
 $N_{w,{\rm pair}}(R)$ is the weighted number of pairs
used in the summation of each radial bin, defined as
%
\beq
N_{w,{\rm pair}}(R)\equiv \left.\sum_{l,s} w_{ls}\right|_{R=\chi_l|\bmtheta_l-\bmtheta_s|}.
\label{eq:Npair_w}
\eeq
Throughout this paper we assume that both redshifts of each lensing halo and each source galaxy
are available via its spectroscopic or photometric redshift.
For the moment we consider a case that the weight depends on lens and source redshifts for simplicity:
$w_{ls}=w(z_l,z_s)$. 
The summation runs over all the pairs each of which has the same projection separation, $R=\chi_l|\bmtheta_l-\bmtheta_s|$, to within the bin width.
The dimension of the stacked lensing profile is $[M_\odot/({\rm Mpc})^2]$.
In the above estimator we multiply the measured ellipticity, $\kappa(\btheta_s,z_s)$, with 
$\sigmacr(z_l,z_s)$ for each pair of source galaxy at $z_s$ and lensing halo at $z_l$
so that the estimator gives an estimation of the average surface mass density profile around the halos, following the method used in the literature
\citep{Mandelbaumetal:13,Miyatakeetal:15}. What is also often used is the stacked lensing estimator without the $\sigmacr$ weight in Eq.~(\ref{eq:est_sigma}), and 
we will discuss the difference below.

In the following, we assume that we select a sample of source galaxies based on their photometric redshifts, $z_s>z_{\rm cut}$, 
where $z_{{\rm cut}}$ is a redshift cut satisfying  $\chi(z_{{\rm cut}})> \chi_l+\Delta \chi_l/2$, i.e. the condition that all the source
galaxies are indeed behind all the lensing halos. 
In this case, the mean number density of all the source galaxies used in the stacked lensing measurement is given as
%
\beq
\bar{n}_{s}\equiv \bar{n}_{\rm tot}\int_{z_{{\rm cut}}}^\infty\!\!\mathrm{d}z_s~ p(z_s).
\label{eq:ns_def}
\eeq
Obviously $\bar{n}_s<\bar{n}_{\rm tot}$.
The ensemble average of the number of the lens-source pairs is given as
%
\begin{align}
\bar{N}_{\rm pair}(R)=&\avrg{\left.\sum_{l,s} w(z_l,z_s)\right|_{R=\chi_l|\bmtheta_l-\bmtheta_s|}}
=\Omega_{S}^2\bar{n}_{\rm tot}\int_{z_{\rm cut}}^\infty\!\!\mathrm{d}z_s~p(z_s)
\int_{\chi_l-\Delta\chi_l/2}^{\chi_l+\Delta\chi_l/2}\!\!\mathrm{d}\chi~\chi^2 f(\chi)
\int\!\!\mathrm{d}M\frac{\mathrm{d}n}{\mathrm{d}M}S(M,\chi) w(z,z_s)\nonumber\\
\simeq& \Omega_S^2\bar{n}_{\rm tot}\int_{z_{\rm cut}}^\infty\!\!\mathrm{d}z_s~p(z_s) w(z_l,z_s)
\int_{\chi_l-\Delta\chi_l/2}^{\chi_l+\Delta\chi_l/2}\!\!\mathrm{d}\chi~\chi^2 f(\chi)
\int\!\!\mathrm{d}M\frac{\mathrm{d}n}{\mathrm{d}M}S(M,\chi) \nonumber\\
=&\Omega_S^2 \bar{n}_s\bar{n}_{\rm h}
\avrg{w(z_l,z_s)}_{z_s},
\end{align}
where $\Omega_S$ is the solid angle of the survey area, and 
$\avrg{w(z_l,z_s)}_{z_s}$ is the averaged weight over the source redshift distribution, defined as
\beq
\avrg{w(z_l,z_s)}_{z_s}\equiv \frac{1}{\int_{z_{\rm cut}}^\infty\!\!\mathrm{d}z_s~p(z_s)}
\int_{z_{\rm cut}}^\infty\!\!\mathrm{d}z_s~p(z_s)w(z_l,z_s),
\eeq
and we have assumed 
that the distribution of lensing halos on the sky 
is uncorrelated with that of source galaxies. We also assumed that 
a geometry of the survey field is sufficiently homogeneous and continuous; in other words, we do not consider 
the effect of survey window for the analytical calculations in this paper.

Using the Limber's approximation
\footnote{\msrev{
Limber's approximation will be violated
when one work on large-scale angular clustering with halos 
within a thin redshift slice.
The validity of Limber's approximation in stacked lensing analyses has been investigated in \citet{2009PhRvD..80l3527J},
while \citet{2017JCAP...11..054A}
validated the Limber's approximation in
computation of halo power spectra.
They showed the accuracy of Limber's approximation is an order of 1\% for multipole larger than 10, 
corresponding to the wavenumber of $\simgt0.01\, h\, {\rm Mpc}^{-1}$
at lens redshift of $z_{\rm lens}=0.3$.}}, 
the ensemble average of the estimator (Eq.~\ref{eq:est_sigma}) is computed as
%
\begin{align}
\avrg{\Sigma}\!(R)=&\frac{1}{\avrg{w(z_l,z_s)}_{z_s}\bar{n}_{{\rm h}}
\int_{z_{\rm cut}}^\infty\!\!\mathrm{d}z_s~p(z_s)}
\int_{z_{\rm cut}}^\infty\!\!\mathrm{d}z_s~p(z_s) \int_{\chi_l-\Delta\chi_l/2}^{\chi_l+\Delta\chi_l/2}\!\!\mathrm{d}\chi~\chi^2 f_{\rm h}(\chi)
\int_0^{\chi_s}\!\!\mathrm{d}\chi' ~ w(z,z_s)
\sigmacr(z,z_s)\sigmacri(z',z_s)\bar{\rho}_{\rm m0}\nonumber\\
&\hspace{6em}\times
\int\!\!\mathrm{d}M\frac{\mathrm{d}n}{\mathrm{d}M}S(M,\chi) \left.\avrg{\delta_{\rm h}(\chi,\chi\bmtheta_l)\delta_{\rm m}(\chi',\chi'\bmtheta_s)}
\right|_{R=\chi|\bmtheta_l-\bmtheta_s|}\nonumber\\
=&\frac{1}{\avrg{w(z_l,z_s)}_{z_s}\bar{n}_{{\rm h}}\int_{z_{\rm cut}}^\infty\!\!\mathrm{d}z_s~p(z_s)}
\int_{z_{\rm cut}}^\infty\!\!\mathrm{d}z_s~p(z_s) \int_{\chi_l-\Delta\chi_l/2}^{\chi_l+\Delta\chi_l/2}\!\!\mathrm{d}\chi~\chi^2f_{\rm h}(\chi)
\int_0^{\chi_s}\!\!\mathrm{d}\chi' ~ w(z,z_s)
\sigmacr(z,z_s)\sigmacri(z',z_s)\bar{\rho}_{\rm m0}\nonumber\\
&\hspace{6em}\times
\int\!\!\mathrm{d}M\frac{\mathrm{d}n}{\mathrm{d}M}S(M,\chi)\left.\int\!\!\frac{\mathrm{d}k_\parallel\mathrm{d}^2\bk_\perp}{(2\pi)^3}
P_{\rm hm}(k;\chi,\chi',M)e^{ik_\parallel(\chi-\chi')+i\bk_\perp\cdot(\chi\bmtheta_l-\chi'\bmtheta_s)}
\right|_{R=\chi|\bmtheta_l-\bmtheta_s|}\nonumber\\
\simeq &\frac{1}{\avrg{w(z_l,z_s)}_{z_s}\bar{n}_{{\rm h}}\int_{z_{\rm cut}}^\infty\!\!\mathrm{d}z_s~p(z_s)}
\int_{z_{\rm cut}}^\infty\!\!\mathrm{d}z_s~p(z_s)w(z_l,z_s) \int_{\chi_l-\Delta\chi_l/2}^{\chi_l+\Delta\chi_l/2}\!\!\mathrm{d}\chi
~ \chi^2f_{\rm h}(\chi)\bar{\rho}_{\rm m0}
\nonumber\\
&\hspace{6em}\times
\int\!\!\mathrm{d}M\frac{\mathrm{d}n}{\mathrm{d}M}S(M,\chi)\left.\int\!\!\frac{\mathrm{d}^2\bk_\perp}{(2\pi)^2}
P_{\rm hm}(k_\perp;\chi, M)e^{i\bk_\perp\chi\cdot(\bmtheta_l-\bmtheta_s)}
\right|_{R=\chi|\bmtheta_l-\bmtheta_s|}\nonumber\\
=&\frac{1}{\bar{n}_{{\rm h}}}\int_{\chi_l-\Delta\chi_l/2}^{\chi_l+\Delta\chi_l/2}\!\!\mathrm{d}\chi
~ \chi^2f_{\rm h}(\chi)\bar{\rho}_{\rm m0}
\int\!\!\mathrm{d}M\frac{\mathrm{d}n}{\mathrm{d}M}S(M,\chi)\int\!\!\frac{k_\perp\mathrm{d}k_\perp}{2\pi}
P_{\rm hm}(k_\perp;\chi,M)J_0(kR).
\end{align}
Thus the stacked lensing for a sample of halos depends only on the matter distribution at the lens redshift, $z_l$. In other words, 
the lensing effects on source galaxies, but at different redshifts from $z_l$ along the line-of-sight, cancel out after the average, because 
the line-of-sight structures are not physically correlated with the distribution of lensing halos.
Hence we can rewrite the projected correlation function in terms of the projected power spectrum as
%
\beq
\avrg{\Sigma}(R)\equiv \int\!\!\frac{k_\perp\mathrm{d}k_\perp}{2\pi}~ C_{\dsigma}(k)J_0(kR), 
\label{eq:sigmaR2}
\eeq
%
where $C_{\dsigma}(k)$ is the projected power spectrum, defined as
%
\beq
C_{\dsigma}(k_\perp)=\frac{1}{\bar{n}^{\rm 2D}_{{\rm h}}}\int_{\chi_l-\Delta\chi_l/2}^{\chi_l+\Delta\chi_l/2}\!\!\mathrm{d}\chi
~ \chi^2f_{\rm h}(\chi)\bar{\rho}_{\rm m0}
\int\!\!\mathrm{d}M\frac{\mathrm{d}n}{\mathrm{d}M}S(M,\chi)
P_{\rm hm}(k_\perp;\chi, M).
\label{eq:C_dsigma}
\eeq
The excess surface mass density profile that can be measured from the stacking of background shapes is similarly expressed in terms of the power spectrum \citep{Hikageetal:13,Murataetal:17} as
%
\beq
\avrg{\dsigma}\!(R)=\int\!\!\frac{k\mathrm{d}k}{2\pi}~C_\dsigma(k)J_2(kR),
\eeq
where $J_2(x)$ is the 2nd-order Bessel function. Thus the stacked lensing estimator, Eq.~(\ref{eq:est_sigma}), is an unbiased estimator in a sense 
that it probes the average matter distribution around lensing halos or the matter-halo cross correlation function at lens redshift, regardless 
of the weight function (the ensemble average does not depend on the weight function). 

In Appendix~\ref{app:cov}, we give a detailed derivation of the 
the covariance matrix for the stacked lensing power spectrum assuming that
both the cosmic shear field and the number density field of halos follow the Gaussian statistics. 
The covariance matrix is expressed as
\begin{equation}
{\cal C}^\dsigma_{ij}\equiv {\rm Cov}[C_{\dsigma}(k_i),C_{\dsigma}(k_j)]=\frac{\delta^K_{ij}}{N_{\rm mode}(k_i)}
\left[C_\dsigma(k_i)^2 + 
\left(C_{\rm hh}(k_i)+\frac{\chi_l^2}{\bar{n}_{\rm h}}\right)
\left(C_{\kappa\kappa,\sigmacr}(k_i)+\frac{\avrg{w(z_l,z_s)^2\sigmacr(z_l,z_s)^2}_{z_s}}{\bar{n}_{s}(\avrg{w(z_l,z_s)}_{z_s})^2}\chi_l^2\sigma_\epsilon^2\right)
\right]
\label{eq:cov_dsigma_main}
\end{equation}
where $\delta^K_{ij}$ is the Kronecker delta function,  and 
\begin{align}
&\hspace{0em}N_{\rm mode}(k_i)\equiv 2 \chi_l^2 f_{\rm sky}k_i\Delta k_i\, ,\nonumber\\
&\hspace{0em}C_{\kappa\kappa,\sigmacr}(k_i)\equiv 
\frac{1}{(\avrg{w(z_l,z_s)}_{z_s})^2}
\int_0^\infty\!\!\mathrm{d}\chi~\avrg{
\sigmacr(z_l,z_s)\sigmacri(z,z_s)w(z_l,z_s)}_{z_s}^2
(\bar{\rho}_{\rm m0})^2
\left(\frac{\chi_l}{\chi}\right)^2
P_{\rm m}\!\left(k=\frac{\chi_l}{\chi}k_i; \chi\right), \label{eq:c_kappakappa_sigmacr}
\end{align}
where $f_{\rm sky}$ is a sky coverage fraction of the survey area, $f_{\rm sky}\equiv \Omega_S/4\pi$, and 
\begin{equation}
\avrg{\sigmacr(z_l,z_s)\sigmacri(z,z_s)w(z_l,z_s)}_{z_s}\equiv 
\frac{1}{\int_{z_{\rm cut}}^\infty\!\!\mathrm{d}z_s~p(z_s)}
\int_{{\rm max}\{z(\chi),z_{\rm cut}\}}^\infty\!\!\mathrm{d}z_s~ p(z_s)\sigmacr(z_l,z_s)\sigmacri(z,z_s)w(z_l,z_s).
\label{eq:kernel_kappakappa_sigmacr}
\end{equation}
Thus we for the first time derived the analytical expression for the covariance matrix of stacked lensing profile, and 
Eq.~(\ref{eq:cov_dsigma_main}) is one of the main results of this paper. 
We should also stress that the covariance formula includes the dependence of weight function.
In the shot noise dominated regime that occur at large $k_i$ bins, the covariance is approximated by
\begin{equation}
{\cal C}^\dsigma_{ij}\simeq \frac{\delta^K_{ij}}{N_{\rm mode}(k_i)}\frac{\chi_l^4}{\bar{n}^{\rm 2D}_{\rm h}}
\frac{\avrg{w(z_l,z_s)^2\sigmacr(z_l,z_s)^2}_{z_s}}{\bar{n}_s(\avrg{w(z_l,z_s)}_{z_s})^2}\sigma_\epsilon^2.
\end{equation}
On the other hand, in the sample variance dominated regime that corresponds to small $k_i$ bins, the covariance
is approximated by 
\begin{equation}
{\cal C}^\dsigma_{ij}\simeq \frac{\delta_{ij}^K}{N_{\rm mode}(k_i)}
\left[C_\dsigma(k_i)^2+C_{\rm hh}(k_i)C_{\kappa\kappa,\sigmacr}(k_i)\right].
\end{equation}

Once the expression of the error covariance matrix is obtained, we can discuss which weight is ``optimal'' to 
maximize the expected signal-to-noise ratio for the stacked lensing measurement for a given survey specification 
and for a 
given cosmological model. The expected signal-to-noise ratio at a given $k$ bin is given as
\begin{equation}
\left(\frac{S}{N}\right)^2_{\dsigma,k_i}\equiv 
\frac{\left[C_\dsigma(k_i)\right]^2}{{\cal C}^\dsigma_{ii}}.
\label{eq:sn_dsigma}
\end{equation}
As we have shown, the covariance matrix ${\cal C}^\dsigma_{ij}$ depends on the weight function, $w(z_l,z_s)$, and therefore 
the signal-to-noise ratio varies with a choice of the weight function. Differentiating $\ln (S/N)^2$ with respect to $w(z_l,z_s)$ and 
setting to zero, we can find an ``optimal'' weight that maximizes the signal-to-noise ratio. For the shot noise limited regime,
the optimal weight is found to be
\begin{equation}
w^{\rm opt}(z_l,z_s)=\frac{\sigmacr^{-2}(z_l,z_s)}{\sigma_\epsilon^2}.
\label{eq:wopt_dsigma}
\end{equation}
This is consistent with what was proposed in \cite{2004AJ....127.2544S, 2005MNRAS.361.1287M, Mandelbaumetal:13}.
Since we do not consider the distribution of intrinsic ellipticities in this paper for simplicity, the optimal weight is 
equivalent to $w^{\rm obt}=\sigmacr^{-2}(z_l,z_s)$. 

For the sample variance limited regime, differentiating $\ln(S/N)^2$ with respect to $w(z_l,z_s)$ and 
setting it to zero, we arrive at 
\begin{align}
&\int_0^\infty\!\!\mathrm{d}\chi~
(\bar{\rho}_{\rm m0})^2\left(\frac{\chi_l}{\chi}\right)^2P_{\rm m}\!\left(k=\frac{\chi_l}{\chi}k_i;\chi\right)
\left[\int_{{\rm max}\{z(\chi),z_{\rm cut}\}}^\infty\!\!\mathrm{d}z_s^{\prime\prime}~p(z_s^{\prime\prime})
\sigmacr(z_l,z_s^{\prime\prime})\sigmacri(z,z_s^{\prime\prime})w(z_l,z_s^{\prime\prime})
\right]\nonumber\\
&\hspace{6em}\times
\left[
\int_{{\rm max}\{z(\chi),z_{\rm cut}\}}^\infty\!\!\mathrm{d}z_s^\prime~ p(z_s^\prime)\sigmacr(z_l,z_s^\prime)\sigmacri(z,z_s^\prime)w(z_l,z_s^\prime)
-\left\{\int_{z_{\rm cut}}^\infty\!\!\mathrm{d}z_s^\prime~p(z_s^\prime)w(z_l,z_s^\prime)\right\}
\sigmacr(z_l,z_s)\sigmacri(z,z_s)
\right] = 0.
\end{align}
%
However, we cannot analytically solve this equation to obtain an expression for the optimal weight, because of complicated 
dependences of 
the cosmic shear contribution on source and lens redshifts and wavenumber. 

\subsection{Stacked shear profile}

An alternative estimator of the stacked lensing profile, often used in the literature \citep[e.g.,][]{Okabeetal:10,PratetalDESgglensing:17},
is 
%
\begin{equation}
\widehat{\avrg{\kappa}}(R)\equiv \left.\frac{1}{N_{w,{\rm pair}}(R)}\sum_{l,s}
w(z_l,z_s)
\kappa(\bmtheta_s,z_s)\right|_{R=\chi_l|\bmtheta_l-\bmtheta_s|}.
\end{equation} 
This estimator is defined without the weight of the critical surface density $\sigmacr(z_l,z_s)$ for each lens-source pair, compared
to Eq.~(\ref{eq:est_sigma}). 
The ensemble average of the above estimator is computed as 
\begin{eqnarray}
\avrg{\kappa}\!(R)&=&
\frac{1}{\avrg{w(z_l,z_s)}_{z_s}
\bar{n}_{\rm h}\int_{z_{\rm cut}}^\infty\!\!\mathrm{d}z_s~p(z_s)}
\int_{z_{\rm cut}}^\infty\!\!\mathrm{d}z_s~p(z_s)\int_{\chi_l-\Delta\chi_l/2}^{\chi_l+\Delta\chi_l/2}\!\!\mathrm{d}\chi~
\chi^2 f_{\rm h}(\chi)
\int_0^{\chi_s}\!\!\mathrm{d}\chi'~w(z_l,z_s)\sigmacri(z',z_s)\bar{\rho}_{\rm m0}\avrg{\delta_{\rm h}(\chi,\chi\bmtheta_l)\delta_{\rm m}(\chi',\chi'\bmtheta_s)}
\nonumber\\
&\simeq&
\frac{1}{\avrg{w(z_l,z_s)}_{z_{\rm cut}}\bar{n}_{\rm h}\int_{z_{\rm cut}}^\infty\!\!\mathrm{d}z_s~p(z_s)}
\int_{\chi_l-\Delta\chi_l/2}^{\chi_l+\Delta\chi_l/2}\!\!\mathrm{d}\chi~\chi^2
f_{\rm h}(\chi)\left[
\int_{{\rm max}\{z_{\rm cut},z(\chi)\}}^\infty\!\!\mathrm{d}z_s~p(z_s)
\sigmacri(z,z_s)\bar{\rho}_{\rm m0}\right]\nonumber\\
&&\hspace{16em}\times \int\!\!\mathrm{d}M\frac{\mathrm{d}n}{\mathrm{d}M}S(M,\chi)
\int\!\!\frac{\mathrm{d}^2\bk_\perp}{(2\pi)}
P_{\rm hm}\left(k_\perp;\chi\right)e^{i\bk_\perp\cdot\chi(\bmtheta_l-\bmtheta_s)}
\nonumber\\
&=&
\frac{1}{\avrg{w(z_l,z_s)}_{z_s}\bar{n}_{\rm h}}
\int_{\chi_l-\Delta\chi_l/2}^{\chi_l+\Delta\chi_l/2}\!\!\mathrm{d}\chi~
\chi^2 f_{\rm h}(\chi)
\avrg{\sigmacri(z,z_s)w(z_l,z_s)
}_{z_s}\bar{\rho}_{\rm m0}
\left(\frac{\chi_l}{\chi}\right)^2\nonumber\\
&&\hspace{16em}\times
\int\!\!\mathrm{d}M\frac{\mathrm{d}n}{\mathrm{d}M}S(M,\chi)
\int\!\!\frac{k_\perp\mathrm{d}k_\perp}{2\pi}
P_{\rm hm}\left(k_{s\perp}=\frac{\chi_l}{\chi}k_\perp;\chi\right)J_0(kR)\nonumber\\
&\simeq&
\frac{1}{\bar{n}_{\rm h}}
\int_{\chi_l-\Delta\chi_l/2}^{\chi_l+\Delta\chi_l/2}\!\!\mathrm{d}\chi~
\chi^2 f_{\rm h}(\chi)
\avrg{\sigmacri(z,z_s)
}_{z_s}\bar{\rho}_{\rm m0}
\left(\frac{\chi_l}{\chi}\right)^2
\int\!\!\mathrm{d}M\frac{\mathrm{d}n}{\mathrm{d}M}S(M,\chi)
\int\!\!\frac{k_\perp\mathrm{d}k_\perp}{2\pi}
P_{\rm hm}\left(k_{s\perp}=\frac{\chi_l}{\chi}k_\perp;\chi\right)J_0(kR),
\end{eqnarray}
where we have assumed the narrow redshift width of lens halos in the last equality, and 
\begin{equation}
\avrg{\sigmacri(z,z_s)}_{z_s}\equiv \frac{1}{\int_{z_{\rm cut}}^\infty\!\!\mathrm{d}z_s~p(z_s)}
\int_{\max\{z(\chi),z_{\rm cut}\}}^\infty\!\!\mathrm{d}z_s~p(z_s) \sigmacri(z,z_s).
\end{equation}
Hence, similarly to Eq.~(\ref{eq:sigmaR2}), we can rewrite 
 $\avrg{\kappa}(R)$ in terms of the projected power spectrum as
%
\begin{equation}
\avrg{\kappa}\!(R)=\int\!\!\frac{k_\perp\mathrm{d}k_\perp}{2\pi}~C_{\gamma_+}(k_\perp)J_0(k_\perp R), 
\end{equation}
where 
\begin{equation}
C_{\gamma_+}(k)\equiv\frac{1}{\avrg{w(z_l,z_s)}_{z_s}\bar{n}_{\rm h}} 
\int_{\chi_l-\Delta\chi_l/2}^{\chi_l+\Delta\chi_l/2}\!\!\mathrm{d}\chi~\avrg{\sigmacri(z,z_s)w(z_l,z_s)}_{z_s}
\bar{\rho}_{\rm m0}\left(\frac{\chi_l}{\chi}\right)^2\int\!\!\mathrm{d}M\frac{\mathrm{d}n}{\mathrm{d}M}S(M,\chi)
P_{\rm hm}\!\!\left(k_{s\perp}=\frac{\chi_l}{\chi}k;\chi\right)
\end{equation}
The stacked shear profile, which is a direct observable from galaxy shapes, is 
\begin{equation}
\avrg{\gamma_+}\!(R)\equiv \int\!\!\frac{k\mathrm{d}k}{2\pi}~C_{\gamma_+}(k)J_2(kR).
\end{equation}

The covariance matrix for the power spectrum $C_{\gamma_+}(k)$ is 
\begin{equation}
{\cal C}^{\gamma_+}_{ij}\equiv {\rm Cov}\left[\hat{C}_{\gamma_+}(k_i),\hat{C}_{\gamma_+}(k_j)\right]
=\frac{\delta^K_{ij}}{N_{\rm mode}(k_i)}
\left[
C_{\gamma_+}(k_i)^2+
\left(C_{\rm hh}(k_i)+\frac{\chi_l^2}{\bar{n}_{\rm h}}\right)
\left(C_{\kappa\kappa}(k_i;z_{\rm cut})+\frac{\avrg{w(z_l,z_s)^2}_{z_s}}{\bar{n}_s(\avrg{w(z_l,z_s)}_{z_s})^2}\chi_l^2\sigma_\epsilon^2\right )
\right]
\end{equation}
where 
\begin{equation}
C_{\kappa\kappa,z_{\rm cut}}(k;z_{\rm cut})\equiv 
\frac{1}{(\avrg{w(z_l,z_s)}_{z_s})^2}\int_0^\infty\!\!\mathrm{d}\chi~ 
\left[
\avrg{\sigmacri(z,z_s)w(z_l,z_s)}_{z_s}
\right]^2
(\bar{\rho}_{\rm m0})^2
\left(\frac{\chi_l}{\chi}\right)^2~
P_{\rm m}\!\left(k_s=\frac{\chi_l}{\chi}k;\chi\right)
\end{equation}
%
%

Similarly to Eq.~(\ref{eq:sn_dsigma}), we can define the expected signal-to-noise ratio for a measurement of 
$C_{\gamma_+}(k)$ from a given survey:
\begin{equation}
\left(\frac{S}{N}\right)_{\gamma_+,k_i}\equiv \frac{C_{\gamma_+}(k_i)^2}{{\cal C}^{\gamma_+}_{ii}}.
\label{eq:sn_gamma}
\end{equation}
For the shot noise dominated regime, the optimal weight that maximizes the signal-to-noise is found to be
\begin{equation}
w^{\rm opt}(z_l,z_s)=\frac{1}{\sigma_\epsilon^2}={\rm constant}.
\label{eq:wopt_gamma}
\end{equation}
If there is no dependence of the intrinsic ellipticity on source redshift, the weight does not change the signal-to-noise ratio, in contrary to the case
for $C_{\dsigma}(k_i)$ (Eq.~\ref{eq:wopt_dsigma}). 

Thus, by defining the stacked shear profile against projected centric radius from lensing halos in the same way, we can compare
the covariance matrices for the two estimators, $\hat{C}_{\dsigma}(k)$ and $\hat{C}_{\gamma_+}(k)$. We can address a question of which estimator 
gives a greater signal-to-noise ratio even if using the exactly same number of lens-source pairs for the same survey area.


\section{Results}
\label{sec:results}
\subsection{Validation of our model with numerical simulations}
\label{subsec:validate}
\subsubsection*{Numerical simulations}

In order to validate our models on estimators of stacked lensing analysis, we utilize a large set of weak gravitational lensing simulations and dark-matter halo catalogs with all sky coverage. 
Here we briefly describe full-sky lensing and halo catalogs, while the details of these catalogs are found in \citet{2017ApJ...850...24T} 
\citep[also see][]{Shirasakietal:17}.
In \citet{2017ApJ...850...24T},
the authors performed a set of $N$-body simulations with $2048^3$ particles 
in cosmological volumes and used them to 
construct lensing and halo catalogs. 
They adopted the standard $\Lambda$CDM cosmology that is consistent with the WMAP cosmology
\citep{2013ApJS..208...19H}.
The cosmological parameters are 
the CDM density parameter 
$\Omega_{\rm cdm}=0.233$, 
the baryon density 
$\Omega_{\rm b}=0.046$, 
the matter density 
$\Omega_{\rm m}=\Omega_{\rm cdm}+\Omega_{\rm b}
= 0.279$, 
the cosmological constant 
$\Omega_{\Lambda}=0.721$, 
the Hubble parameter
$h= 0.7$, 
the amplitude of density fluctuations
$\sigma_8= 0.82$,
and the spectral index
$n_s= 0.97$.
In the following,
we 
use 
10 full-sky lensing simulations that are selected from 
108 realizations in \citet{2017ApJ...850...24T}.

\subsubsection*{Lensing catalog}

Full-sky weak gravitational lensing simulations
have been performed with the standard multiple lens-plane algorithm \citep[e.g.][]{2001MNRAS.327..169H, 2013MNRAS.435..115B, 2015MNRAS.453.3043S}. 
In this simulation, one can take into account the light-ray deflection on the celestial sphere
by using the projected matter density
field given in the format of spherical shell.
The simulations used the projected matter fields in 38 shells in total, each of which was computed by projecting $N$-body simulation realization over a radial width of $150\, h^{-1}{\rm Mpc}$, in order to make the light cone covering a cosmological volume up to $z=5.3$.
As a result, the lensing simulations
consist of shear field at 38 different source redshifts with angular resolution of 0.43 arcmin.
Each simulation data is given in the {\tt HEALPix}
format \citep{2005ApJ...622..759G}.
The interval between nearest source redshifts 
is set to be $150\, h^{-1}{\rm Mpc}$ in comoving distance, corresponding to the redshift depth of $0.05-0.1$ for $z\simlt1$.

Using the above lensing simulations, we 
create mock shear catalogs for 
two different hypothetical surveys, 
DES and subaru HSC.
For a given source distribution $p(z_s)$,
we discretize $p(z_s)$ so as to match 
the redshift width in the lensing simulations.
Then, we distribute mock source galaxies according
to given $p(z_s)$ and source number density.
Here we ignore the clustering of source galaxies
and assume random distribution of sources on the sky. For DES-type, we set $\langle z_s \rangle=0.7$
with source number density of 
$7 \, {\rm arcmin}^{-2}$,
while we assume $\langle z_s \rangle=1.0$ with source number density of $20 \, {\rm arcmin}^{-2}$ in HSC-type survey.
For each source galaxy, we assign the intrinsic shape noise by following Gaussian distribution 
with the rms of $\sigma_{\epsilon}=0.3$.

\subsubsection*{Halo catalog}

In each output of the $N$-body simulation, 
\citet{2017ApJ...850...24T} locate dark matter
halos using the {\tt Rockstar} algorithm \citep{2013ApJ...762..109B}. 
Throughout this paper, we define the halo mass by using the spherical overdensity criterion:
$M_{\rm 200m} = 200\bar{\rho}_{\rm m0}(4\pi/3)R^{3}_{\rm 200m}$. 
Individual halos in $N$-body boxes 
are assigned to the pixels in the celestial sphere
with the {\tt HEALPix} software. 
In the following, we consider 
a mass-limited sample with 
$M_{\rm 200m}\ge10^{13.5}\, h^{-1}M_{\odot}$
at redshift of $0.45-0.55$. 
Note that 
the halo with mass of $M_{\rm 200m}=10^{13.5}\, h^{-1}M_{\odot}$
is revolved by $\sim1000$ $N$-body particles
in this redshift range.

\subsubsection*{Mock stacked lensing analysis}

Using mock catalogs that we described above, 
we evaluate the covariance matrices of two different estimators, surface mass density
$\Delta \Sigma$ and lensing shear $\gamma_+$
around foreground halos.
In order to increase the number of realizations,
we divide a full sky into 192 subregions 
with the survey area of $4\pi/192$ str,
corresponding to $215\, {\rm deg}^2$.
Since we work with 10 full-sky realizations, 
we use 1920 realizations of mock shear and foreground halo catalogs 
in stacked lensing analyses in total.
These 1920 realizations allow 
us to evaluate the covariance matrices 
of $\Delta \Sigma$ and $\gamma_+$ 
for DES and HSC-type surveys 
with the sky coverage of $215\, {\rm deg}^2$.
Note that we can find $\sim2500$ halos 
with $M_{\rm 200m}\ge10^{13.5}\, h^{-1}\, M_{\odot}$ at $z_l=0.45-0.55$ in each sky coverage.
\ms{The size of this survey window corresponds to the transverse distance at lens redshift, 
$\chi(z_l)\times \sqrt{\Omega_{\rm S}}\simeq 341~{\rm Mpc}/h$ for the lens redshift $z_l=0.5$.}
For simplicity, we use the all source galaxies behind foreground objects, corresponding 
to $z_{\rm cut}=0.55$.
Applying the criteria of $z_{\rm cut}=0.55$, 
we find the effective source number densities
are equivalent to 
$4.13\, {\rm arcmin}^{-2}$
in DES-type survey and
$15.6\, {\rm arcmin}^{-2}$
in HSC-type survey, respectively.
We correct for the observed signals 
in our mock analysis
by subtracting the stacked lensing around random points. For this subtraction, we set the number of random points to be 10 times as large as one 
of foreground halos.
In mock stacked lensing analysis,
we employ 17 equally-spaced logarithmic bins with the bin width of $\Delta \ln R=0.2$ 
in the range of 
$3.0 < R\, [h^{-1}\, {\rm Mpc}]<73.6$.

\subsubsection*{Comparison of model and simulations}

In this section, we use the mock catalogs of background galaxy shear and foreground halos to validate the covariance matrix for stacked lensing
estimators. 
For comparison, we define the signal-to-noise ratio for stacked lensing profile as $(S/N)^2[{\cal F}] = \langle{\cal F}\rangle^2/{\rm Var}[{\cal F}]$ where ${\cal F}=\Delta \Sigma$ or $\gamma_+$.

The signal and covariance of stacked surface mass density profile in real space 
can be expressed as \citep[also see e.g.][]{OguriTakada:11}
\begin{align}
&\langle\Delta \Sigma\rangle (R_{m}) = \int \frac{{\rm d}k\, k}{2\pi}\, C_{\Delta \Sigma}(k)\, \hat{J}_2 (kR_{m}), \label{eq:dsigma_real}\\
&{\rm Cov}\left[\langle\Delta \Sigma\rangle (R_{m}), \langle\Delta \Sigma\rangle (R_{n})\right] = 
\frac{1}{\Omega_{S}\chi^2_{l}}
\int \frac{{\rm d}k\, k}{2\pi}\, \left[C^2_{\Delta \Sigma}(k) + C^{\rm obs}_{\rm hh}(k)C^{\rm obs}_{\kappa\kappa, \sigmacr}(k)\right]
\hat{J}_2 (kR_{m}) \hat{J}_2 (kR_{n}), \label{eq:cov_dsigma_real}
\end{align}
where 
$C^{\rm obs}_{\rm hh}$ and $C^{\rm obs}_{\kappa\kappa, \sigmacr}$ are defined in Appendix~\ref{app:cov}.
In Eqs.~(\ref{eq:dsigma_real}) and (\ref{eq:cov_dsigma_real}), 
$\hat{J}_{2}(k R_{m})$ is the 2nd-order Bessel function averaged 
within an annulus between $R_{m, \rm min}$ and $R_{m,\rm max}$,
\begin{equation}
\hat{J}_{2}(kR_{m}) = \frac{2}{R^2_{m,\rm max}-R^2_{m, \rm min}}\int_{R_{m,\rm min}}^{R_{m,\rm max}}{\rm d}R\,R\,J_2(kR).
\end{equation}
For the signal and covariance of stacked shear profile in real space, we use the similar expressions as in
Eqs.~(\ref{eq:dsigma_real}) and (\ref{eq:cov_dsigma_real}).
When computing analytic prediction, we use the fitting formula of non-linear matter power spectrum 
$P_m(k,z)$ developed in \citet{2012ApJ...761..152T}.
In addition, we consider the standard halo-model approach
to predict the cross power spectrum between $P_{\rm hm}(k,z)$
and projected halo power spectrum $C^{\rm obs}_{\rm hh}(k)$.
We adopt the model of halo mass function and linear bias in 
\citet{2008ApJ...688..709T, 2010ApJ...724..878T}.
Furthermore, we assume the NFW profile \citep{1997ApJ...490..493N} with the mass-concentration relation as in \citet{2015ApJ...799..108D}.

\begin{figure*}
\centering
\includegraphics[width=0.45\columnwidth]{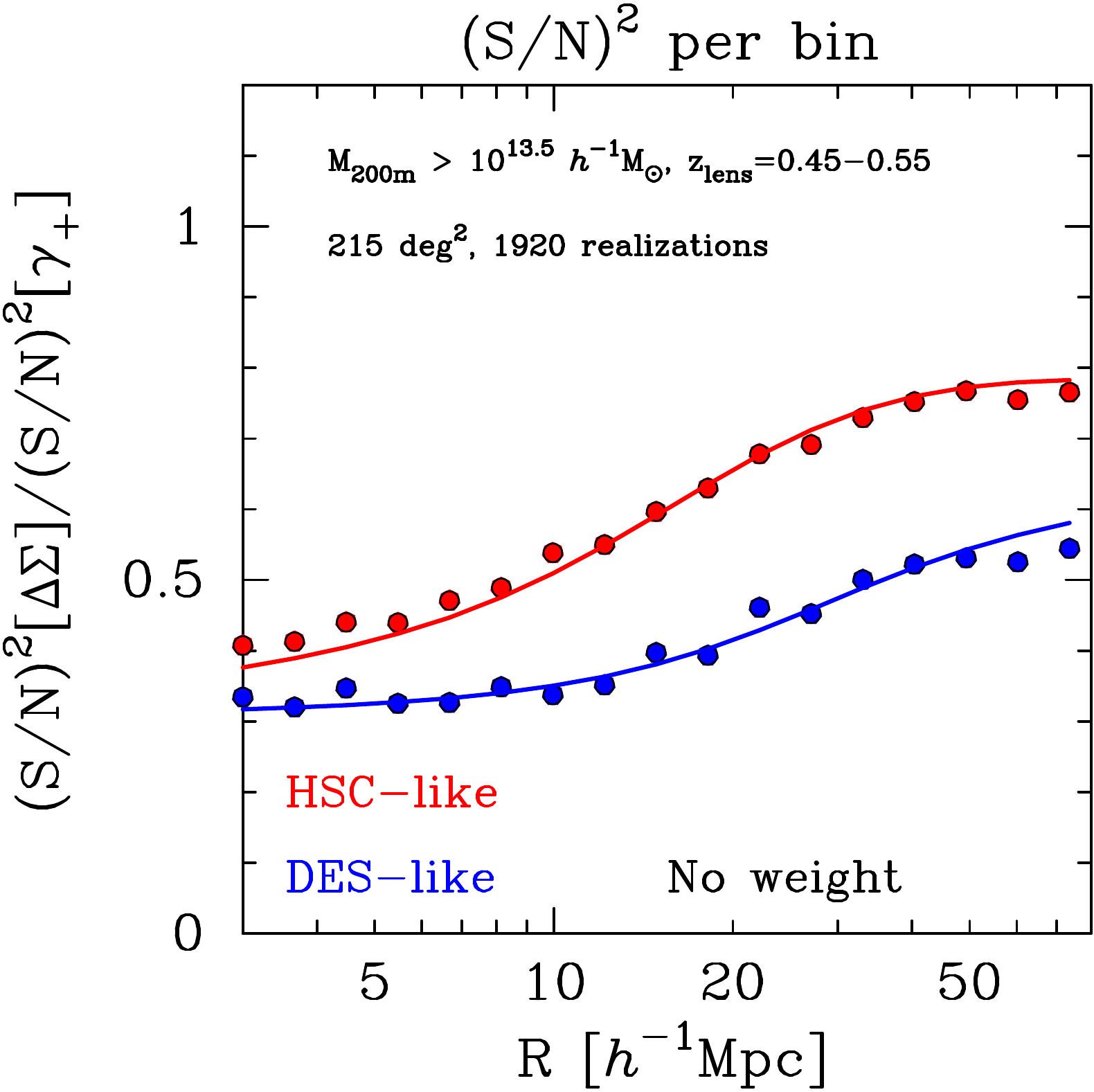}
\includegraphics[width=0.45\columnwidth]{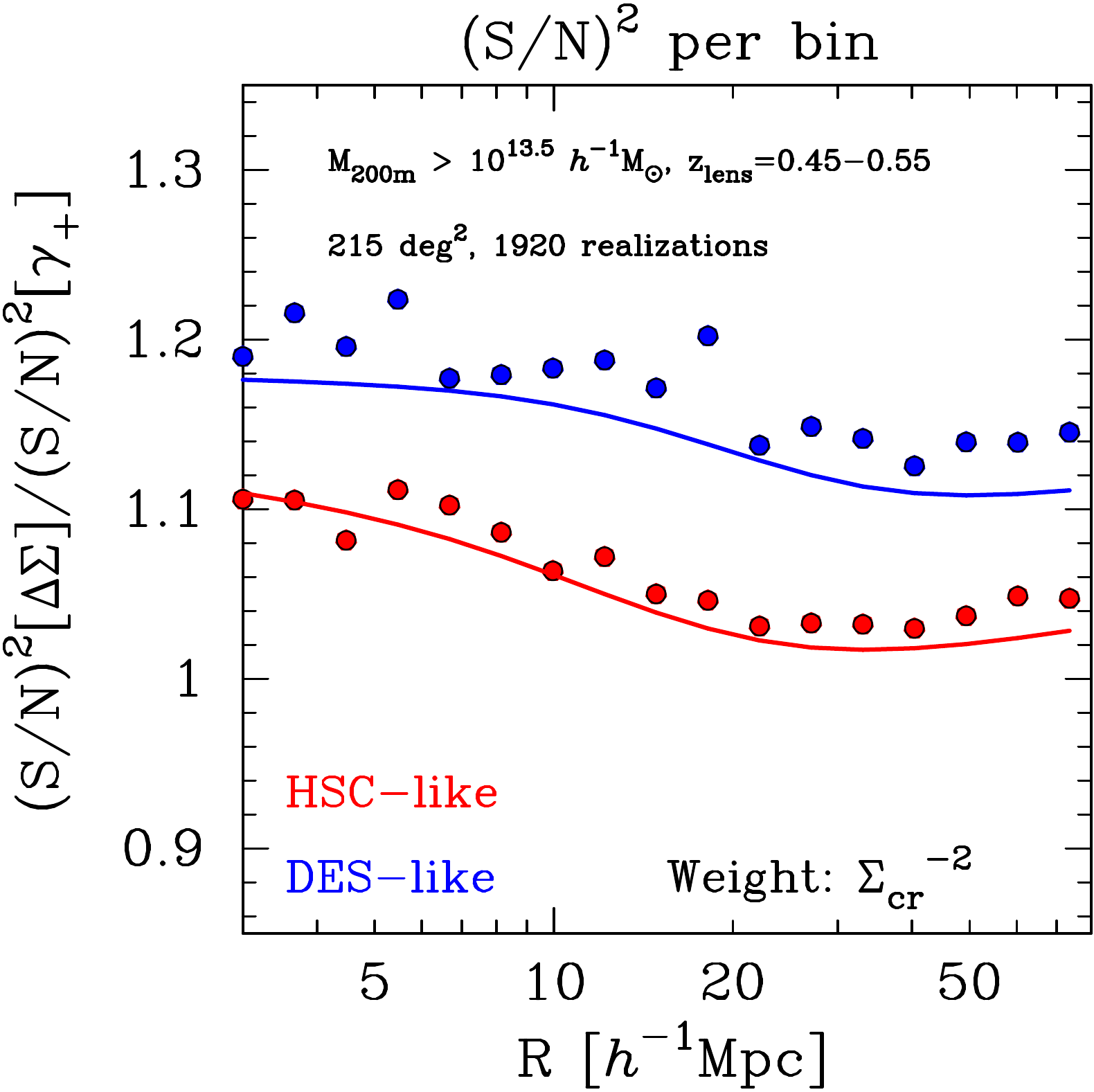}
\caption{
	Comparison of the halo model predictions with the results using mock simulations of stacked lensing. 
    Each panel shows the ratio of $(S/N)^2$ between stacked surface
    mass density $\Delta \Sigma$ and stacked shear profile $\gamma_+$ in each radial bin, $R$. 
    The colored points represent the simulation results, while
    the lines are for the analytical model predictions. 
    The color difference presents different survey parameters.
    In each panel, the red corresponds to HSC-like survey
    and the blue is for DES-like survey (see text for the details).
    {\it Left}: The weight in surface mass density is set to be 1.
    {\it Right}: Similar to the left, but the results if we employ 
    the weight  $\sigmacr^{-2}(z_l, z_s)$ for the signal-to-noise ratios of 
    $\Delta\Sigma$.
     \ms{Note that the right panels show narrower range of the ratio of $(S/N)^2$ than the left.}
	}
\label{fig:model_vs_sims}
\end{figure*}

Figure~\ref{fig:model_vs_sims} shows
the ratio of $(S/N)^2[{\cal F}]$ between two different estimators, $\Delta \Sigma$ and $\gamma_+$.
In this figure, we consider two different weight functions in the $\Delta \Sigma$-estimator.
One is the weight $w(z_l, z_s)$ to be unity
and corresponds to the absence of any weight in the $\Delta\Sigma$-estimator.
Another case assumes $w(z_l, z_s) = \sigmacr^{-2}(z_l, z_s)$ that is the optimal weighting to realize the minimum variance in shot-noise dominated regime.
The left panel of Figure~\ref{fig:model_vs_sims}
presents the case of $w(z_l, z_s)=1$,
while the right is for 
$w(z_l, z_s) = \sigmacr^{-2}(z_l, z_s)$.
The colored points represent the simulation results
and the line corresponds to our analytic predictions.
As seen in Figure~\ref{fig:model_vs_sims},
our theoretical model of different stacked lensing profiles is in reasonably good agreement with
mock simulation results \ms{(within a 2-3$\%$ level)}, for different survey parameters over the range of radii.
The agreement implies that the non-Gaussian error contributions, which should exist in the simulations, are not 
significant for the covariance matrix of stacked lensing profile \ms{(also see Takahashi et al. in prep.).}
The left panel clearly shows that, if the weight for the $\Delta\Sigma$-estimator is not adopted, i.e. if $w(z_l,z_s)=1$, 
the $(S/N)^2$ for $\Delta\Sigma$ is always lower than that for the $\gamma_+$-estimator. 
The right panel shows that, if one adopts 
the weight $w(z_l, z_s) = \sigmacr^{-2}(z_l, z_s)$ for $\Delta\Sigma$, which downweights
source-lens pairs that are close in redshift,
the $S/N$ is improved, because the weight efficiently reduces the shot noise contamination to the measurement. 
As can be found, the use of the weight leads to an improvement in $(S/N)^2$ for the $\Delta\Sigma$-estimator 
compared to that for $\gamma_+$-estimator, at about 5--25\% level depending on survey parameters. 
The improvement is greater for the DES-type survey than in the HSC-type survey, because the median of source redshifts 
in the DES-type survey is closer to lens redshift, $z_l\simeq 0.5$ in this particular case, and the use of the weight 
more efficiently downweights the lensing contributions arising from lens-source pairs that are close in redshift. 

\subsection{Dependence on LSS tracers and survey parameters}

We then study how the cumulative $(S/N)^2$ values for the stacked lensing estimators ($\Delta\Sigma$ vs. $\gamma_+$) 
vary with different combinations of 
lens and source redshifts or/and  mass ranges of lensing halos.
To do this we use the halo model predictions for the stacked lensing in Fourier space, 
because it should contain equivalent information to that for  
the real-space stacked lensing profiles \ms{at least for scales (large $k$) that are much smaller than a size of the survey window.} 
\ms{If one consider a more complex survey window, e.g. with small-scale masks such as masks due to bright stars, 
it causes additional mode-coupling between different Fourier modes, which need to be properly taken into account. However we think that
the following results for the ``relative'' comparison of the two estimators of $\Delta\Sigma$ and $\gamma_+$ still hold valid for a general survey window, because both the estimators are affected by the survey window in the same way.}
We here consider a mass-limited sample of halos with masses greater than a given mass threshold 
$M_{\rm h, min}$ for different lens redshift ranges
of $z_l=[z_{l,\rm min}, z_{l,\rm max}]$.
In this section, we set five different mass bins of 
$\log(M_{\rm h, min}/h^{-1}\, M_{\odot})=13.0, 13.5, 14.0, 14.5$ and 15.0
and five redshift bins of $z_{l,\rm min}=0.05, 0.25, 0.45, 0.65$ and 0.85 with $z_{l,\rm max} = z_{l,\rm min}+0.10$.

\begin{figure*}
\centering
\includegraphics[width=0.45\columnwidth]{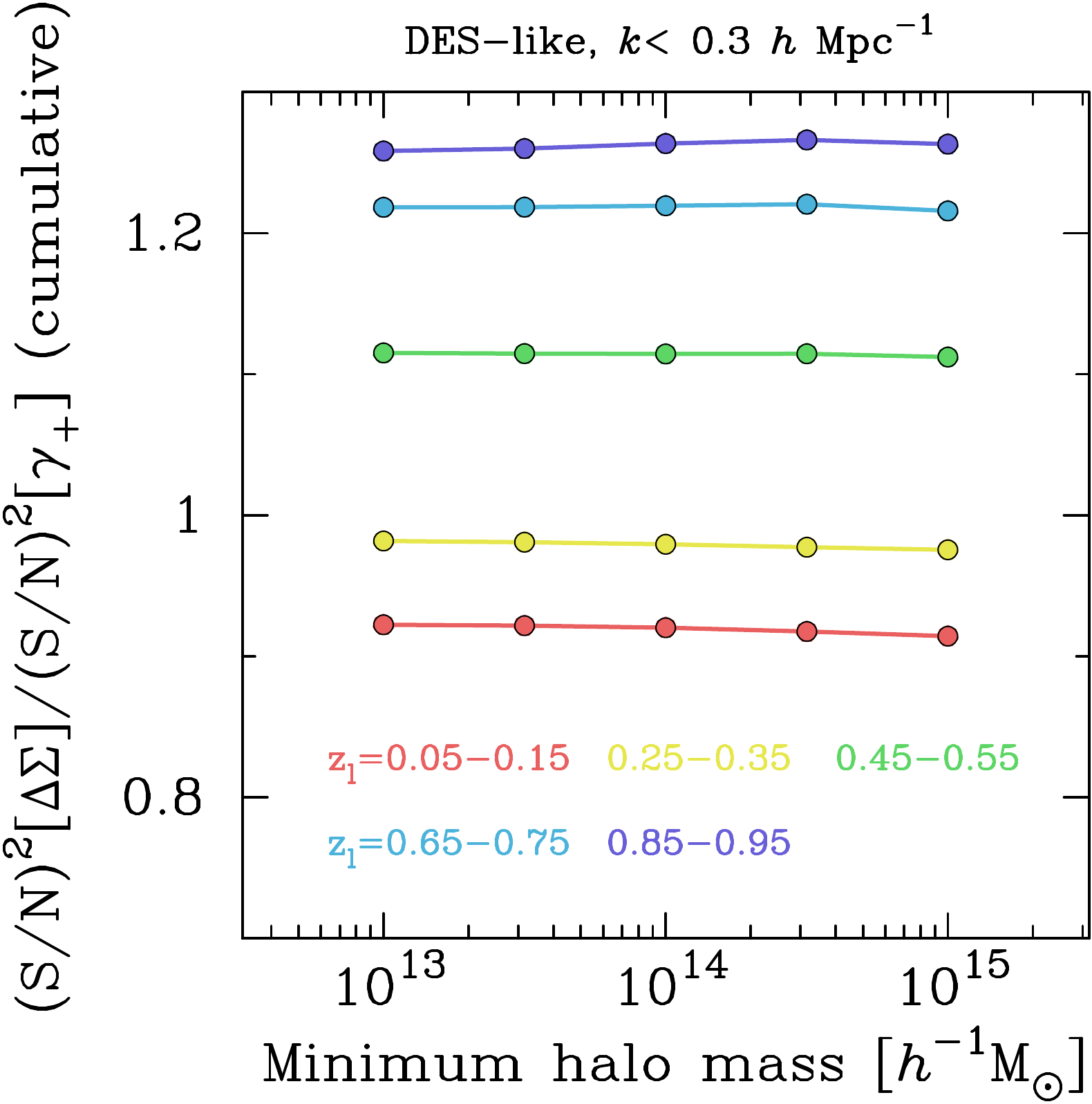}
\includegraphics[width=0.41\columnwidth]{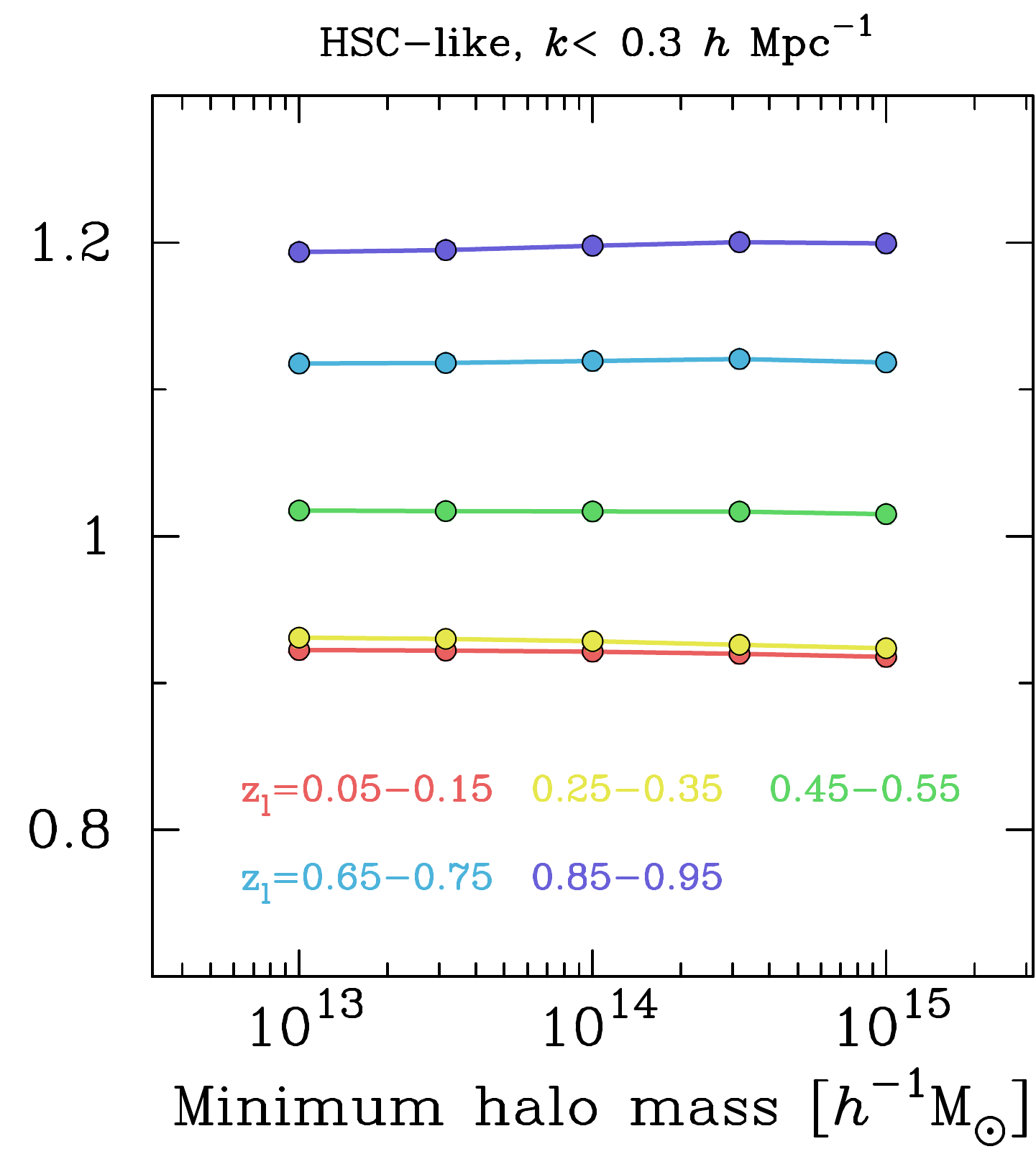}
\caption{
	An improvement or degradation in $(S/N)^2$ if using the weight  $\sigmacr^{-2}$
	for the $\Delta\Sigma$-estimator for 
	a hypothetical measurement of the 
	stacked lensing,
    compared to the $\gamma_+$-estimator. 
    In each panel, we compute the cumulative $(S/N)^2$ by integrating the information of their respective power spectra 
    over the range, $10^{-4}\le k/[h\, {\rm Mpc}^{-1}]\le 0.3$. 
    We show the ratio of $(S/N)^2$ as a function of minimum halo mass
    of foreground halos used in the stacked lensing analysis.
    Different colored lines represent the results for different 
    lens redshifts $z_l$.
    The left assumes DES-like survey with mean source redshift of 0.7
    and source number density of $7\, {\rm arcmin}^{-2}$, while
    the right is for HSC-like survey with mean source redshift of 1.0
    and source number density of $20\, {\rm arcmin}^{-2}$.
	}
\label{fig:sn_ratio_varying_m_z_survey}
\end{figure*}

Figure~\ref{fig:sn_ratio_varying_m_z_survey} shows the ratio of  cumulative $(S/N)^2$ for the two stacked lensing estimators, 
$\Delta\Sigma$ vs. $\gamma_+$. To compute the cumulative $(S/N)^2$ we include the power spectrum information over 
the range of wavenumbers, $10^{-4}\le k/[h\,{\rm Mpc}^{-1}]\le 0.3$, 
define the source galaxies by using $z_{\rm cut} = z_{l,\rm max}$, and employ the weight $\sigmacr^{-2}$ for the $\Delta\Sigma$-estimator as default.
The left- and right-panels show the results for 
 the DES- and HSC-type surveys, respectively.  
In each panel, the colored lines represent the ratio as a function of $M_{\rm h, min}$
and different colors correspond to different lens redshift ranges.
The signal-to-noise ratios for $\Delta\Sigma(R)$ and $\gamma_+(R)$ are 
generally different. However, the difference
is fairly insensitive to the minimum halo mass, but rather sensitive to the lens redshifts relative to source redshifts.
In particular,
we find an improvement in $(S/N)^2$ for the $\Delta\Sigma$-estimator over the $\gamma_+$-estimator, 
when lens redshifts are relatively high. 
To be more precise, a greater improvement can be obtained as lens redshifts increase and 
a 20\%-level improvement can be realized for lens redshift of $z_l\sim0.9$ for the DES- and HSC-type surveys. 
This result indicates that the $\Delta \Sigma$-estimator with the weight  $\sigmacr^{-2}$
is beneficial to extract the greater information in a given survey, especially for stacked lensing measurements
of
galaxy groups and clusters at high redshifts, which
can be obtained in ongoing imaging surveys \citep[e.g.][]{2018PASJ...70S..20O}
and ground-based CMB experiments via the Sunyaev-Zel'dovich effect
\citep[e.g.][]{2013JCAP...07..008H, 2015ApJS..216...27B}.
Contrarily, if lens redshifts are low such as $z_l\simlt 0.3$ compared 
to the median of source redshifts ($\sim0.5$ and 
$\sim0.67$ for DES- and HSC-type surveys in this figure), the $\Delta\Sigma$-estimator is not necessarily 
optimal and does not bring a gain in the stacked lensing measurement. For such low-redshift lenses, one should instead
use the $\gamma_+$-estimator to extract the maximum information. 

\subsection{Increasing the signal-to-noise ratio at sample-variance limited regime}

We were not able to find a unique solution of the weight that can maximize the signal-to-noise ratio 
of stacked lensing
in the sample-variance limited regime. 
Ongoing and future wide-area galaxy surveys will allow us to 
measure the stacked lensing signals up to greater radii such as BAO scales around $R\simeq 100{\rm Mpc}/h$,
at a high significance \citep[e.g.][]{2009PhRvD..80l3527J,dePutterTakada:10}.
The stacked lensing signals at such large radii should include cleaner information on cosmology because 
such signals are still in the weakly nonlinear or linear regimes, are less affected by baryonic physics and are relatively easier to 
model, e.g. based on linear theory or perturbation theory of large-scale structure formation
\citep{OguriTakada:11}.
Hence, it is worth to explore an effective weight that can improve the signal-to-noise ratios 
in the sample variance limited regime.

\begin{figure*}
\centering
\includegraphics[width=0.45\columnwidth]{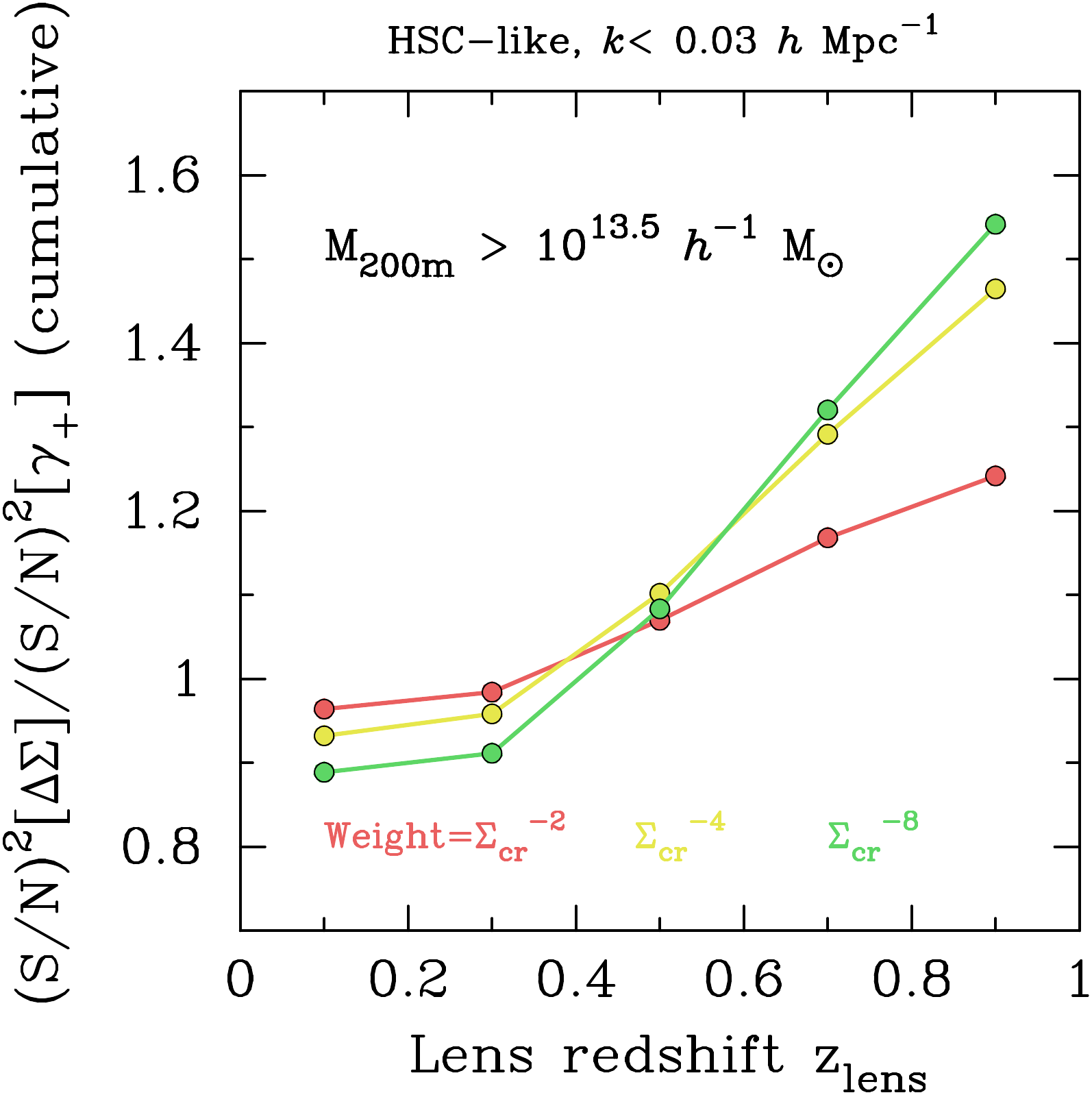}
\includegraphics[width=0.45\columnwidth]{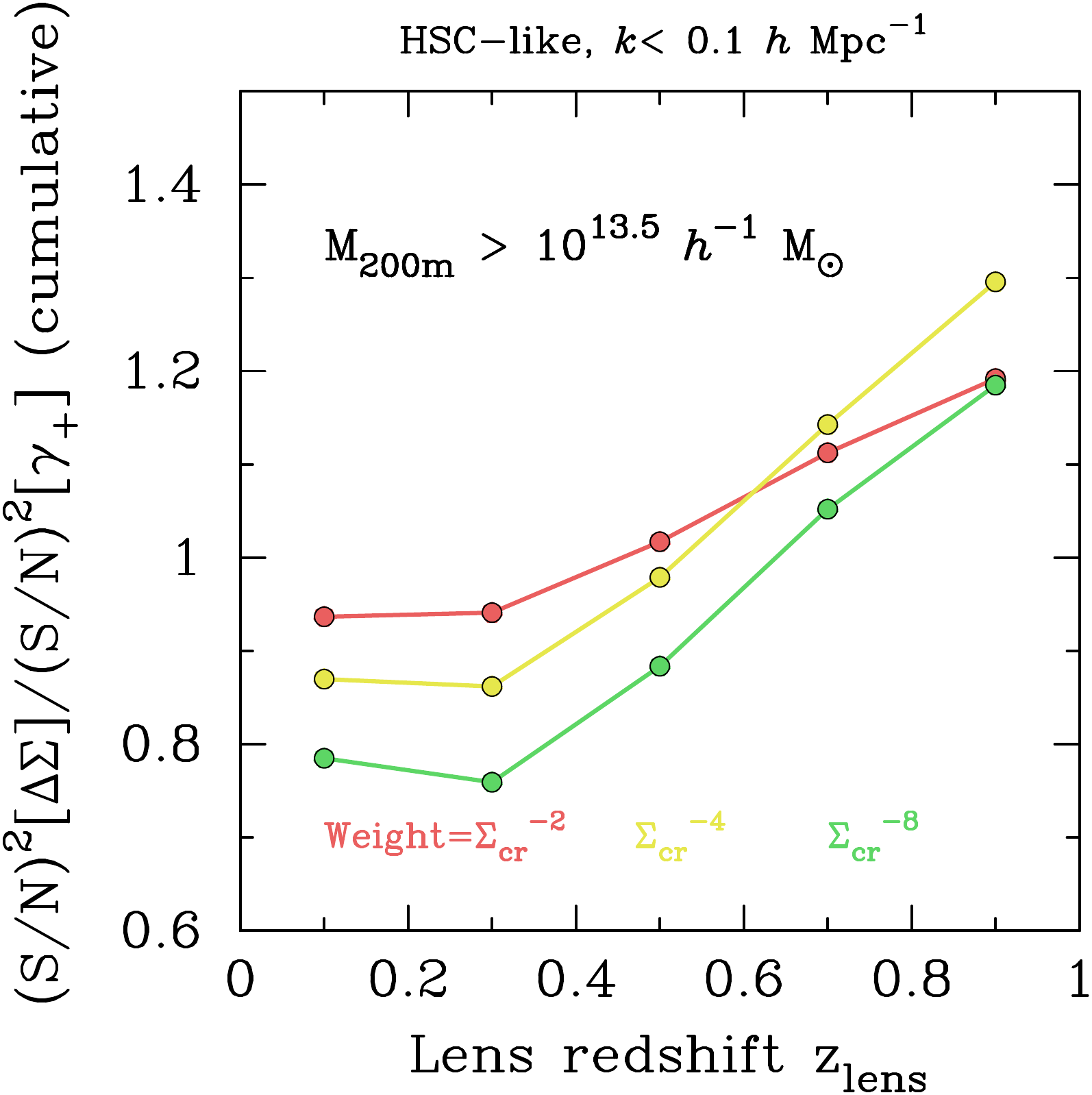}
\caption{An improvement or degradation in the cumulative $(S/N)^2$ in the sample variance limited regime
    by using different weights, parameterized by 
    $w(z_l,z_s)=\sigmacr^{-\alpha}(z_l,z_s)$, for the $\Delta\Sigma$-estimator,
    compared to that for $\gamma_+$.
	We compute the cumulative $(S/N)^2$ for the maximum wavenumber, $k_{\rm max}=0.03$ or $0.1~h\,{\rm Mpc}^{-1}$ up to which
	we include the power spectrum information for the respective estimators for the HSC-type survey, in the left or right panel, respectively. 
	Note that we fixed the minimum wavenumber to $k_{\rm min}=10^{-4}~h\, {\rm Mpc}^{-1}$ for both the cases, and 
	we do not employ any weight (i.e. $w=1$) for the $\gamma_+$-estimator.
    We show the ratio of $(S/N)^2$ as a function of 
    lens redshifts assuming 
    the mass-limited sample with $M>10^{13.5}\, h^{-1}\,M_{\odot}$.
    Different colored lines represent 
    the results with different power-law 
    index in weight function.
	}
\label{fig:LSST_largeR}
\end{figure*}

To address the above question, we here employ an empirical approach as follows.  
From Eq.~(\ref{eq:c_kappakappa_sigmacr}), we expect that, if we further down-weight source galaxies that are closer to lensing halos in redshifts 
or equivalently if we more aggressively up-weight source galaxies that are in higher redshifts,
we could further suppress sample variance contamination that arises from large-scale structure (cosmic shear) at lower redshifts. Since 
large-scale structure is more evolving at lower redshifts, such a weight could help to reduce the statistical scatters in the stacked 
lensing measurements. 
Motivated by this fact, we here consider 
the
weight  $\sigmacr^{-\alpha}$ with different power-law indices
$\alpha=2, 4$ or 8, respectively.
Figure~\ref{fig:LSST_largeR} shows 
the ratio of cumulative $(S/N)^2$ of stacked lensing in Fourier space for the mass-limited sample with $M>10^{13.5}\, h^{-1}\, M_{\odot}$.
Here we assume the HSC-type 
survey as in the previous figures, and we assume $k_{\rm max}=0.03$ or $0.1\, h\, {\rm Mpc}^{-1}$ for the maximum wavenumber up to which we 
include the power spectrum information to compute the cumulative $(S/N)^2$ as in Figure~\ref{fig:sn_ratio_varying_m_z_survey}
in the left or right panel, respectively.
Note that we set $k_{\rm min}=10^{-4}~h\, {\rm Mpc}^{-1}$ for both the cases.  
\ms{Using two different values in $k_{\rm max}$, we study how the improvement of $(S/N)^2$ in sample variance dominated regime 
can depend on the maximum wavenumber in stacked lensing analysis.}
Red, yellow and green lines represent
the ratio of $(S/N)^2$ for $\alpha=2, 4$
and 8, respectively.
Surprisingly, this simple approach using the weight 
of $\sigmacr^{-\alpha}$ with $\alpha>2$ 
is found to be very efficient for improving
the signal-to-noise ratio of stacked lensing profiles.
The improvement is greater for lensing halos at higher redshifts and 
can be up to a factor of 1.5,
which is equivalent to a larger-area survey by the same factor. 
We also found a similar-level improvement for the DES-type survey, but the exact amount of improvement is different due to different relative 
contributions of cosmic shear and shot noise in the covariance elements. 
As we show in Appendix~\ref{app:mock_var_largescale}, we confirm that such an reduction in the sample variance of $\Delta\Sigma$-estimator
is found from the simulations if using the weight $\sigmacr^{-\alpha}$ with $\alpha>2$. 
This method might be useful to obtain an optimal constraint on cosmological parameters such as
the BAO scale and the primordial non-Gaussianity from the large-scale lensing information.
However, it should be noted
that we here do not include
any systematic effects in source redshift estimation.
Increasing a power-law index $\alpha$ in the weight  $\sigmacr^{-\alpha}$
means
that source galaxies at higher redshift are more aggressively up-weighted
in the stacked lensing measurements. 
Therefore, systematic errors in source redshifts could cause a severe systematic bias 
in the measured lensing signals, which then causes a bias in cosmological parameters. 
Even for systematic errors due to source redshift uncertainty, we could marginalize over
the effect by using the method developed in \citet{OguriTakada:11}, which is using a single population of 
source galaxies for lensing objects at multiple redshifts
and then using the redshift dependence of lensing effects 
in the multiple lens planes to marginalize over the systematic error due to source redshift uncertainty. 
This is an interesting possibility and worth exploring.

\section{Conclusions}
\label{sec:conclusions}

In this paper, we derived
a formula for the covariance matrix of stacked lensing profiles, where 
the expression includes
an arbitrary weight function that is given as a function of lens and source redshifts. 
Using the formula, we examined two estimators of stacked lensing used in the literature,
referred to as 
stacked surface mass density profile ($\Delta\Sigma$)
and stacked shear profile ($\gamma_+$).
The former is known as an estimator of the average projected matter density profile around foreground objects,
while the latter characterizes the average gravitational lensing effects induced by foreground objects.
We paid particular attention to the following questions. 
Are the two estimators equivalent in terms of the signal-to-noise ratio if measured from the exactly 
same datasets
(the same
lens-source pairs and the same survey region)? 
How does the use of a weight factor, such as $\sigmacr^{-2}$ for the $\Delta\Sigma$-estimator commonly used in literature, 
improve or degrade the signal-to-noise ratio for the stacked lensing measurement? 
Our findings are summarized as follows:

\begin{enumerate}

\vspace{2mm}
\item 
We showed that the use of the weight $\sigmacr^{-2}$ for the $\Delta\Sigma$-estimator 
is optimal in the shot-noise limited regime since it maximizes the signal-to-noise ratio for a measurement of $\Delta \Sigma$.
This is a different derivation from that in \cite{BernsteinJarvis:02} \citep[also see][]{2004AJ....127.2544S}, which is based on a simple inverse variance weighting to observed ellipticities of source galaxies.
In this paper we did not consider the distribution of intrinsic ellipticities, but it is straightforward to include the effect of the intrinsic 
ellipticities in the weight. On the other hand, in the sample variance limited regime, 
we were not able to find a simple expression for the optimal weight.

\vspace{2mm}
\item 
Using a large set of mock catalogs including shear of background galaxies and lensing halos in the light-cone realization,  
we validated our formula for the 
covariance matrix of stacked lensing profile by comparing the theoretical prediction with the simulation results. 
The use of the weight $\sigmacr^{-2}$, which  
downweights source-lens pairs that are close in redshift, induces
a $\sim5$--$25\%$ improvement 
of $(S/N)^2$ for the $\Delta \Sigma$-estimator, compared to the stacked shear profile $\gamma_+$, 
for DES- and HSC-type surveys, if the redshift of lensing objects is comparable with or higher than the median of background galaxy redshifts. 
This improvement is equivalent to a $\sim 5$--$25$\% larger 
survey area. On the other hand, for low-redshift lenses such as $z_l\simlt 0.3$, the $(S/N)$ for the $\Delta\Sigma$-estimator is lower than
that for the $\gamma_+$-estimator.
For the improvement in the $\Delta\Sigma$-estimator, 
a selection of lensing objects such as lensing halos above a certain mass threshold
is found to be irrelevant.
Hence it is important to 
employ the
weight $\sigmacr^{-2}$ in the $\Delta\Sigma$ 
measurements
for galaxy groups and clusters at high redshifts such as $z_l\simgt 1$, 
which can be found from upcoming wide- and deep-area optical surveys as well as CMB experiments via
the Sunyaev-Zel'dovich effect. 

\vspace{2mm}
\item
We found that 
it is still possible to reduce the statistical uncertainty due to cosmic shear in stacked lensing measurements 
in the sample variance limited regime, by employing the weight 
$\sigmacr^{-\alpha}$ with $\alpha>2$, which more up-weights source galaxies at higher redshifts than the weight $\sigmacr^{-2}$ does.
This is because large-scale structure, which causes the cosmic shear contamination to the sample variance of stacked lensing, 
is more evolving at lower redshifts, and such a weight can suppress the contribution arising from low redshifts (or more aggressively 
up-weight the lensing signals that are from source galaxies at higher redshifts).
We examined the cases with $\alpha=4$ and 8 by using our analytic expressions.
We found that the cumulative $(S/N)^2$ in the power spectrum of $\Delta\Sigma$
up to $k<0.03$ or $0.1\, h\, {\rm Mpc}^{-1}$ 
can be improved by up to  50\%  for the HSC-type survey.  
We also confirmed the reduction of the statistical error in large-scale $\Delta \Sigma$ with $\alpha=4$ and 8 in numerical simulations.

\end{enumerate}

Combining the above results, we conclude that we can employ a hybrid method using different weights 
for the stacked lensing measurements in different regimes; we can employ 
the weight $\sigmacr^{-2}$ for lens-source pairs with small separations that are 
in the shot noise limited regime, while we employ the weight $\sigmacr^{-\alpha}$ with 
$\alpha>2$ for source-lens pairs with large separations that are in the sample variance-limited regime. 
For low-redshift lenses, we should use the $\gamma_+$-estimator rather than the $\Delta\Sigma$-estimator. 
Such an optimal estimator of the stacked lensing
would allow us to extract 
a maximum information of stacked lensing profiles, which in turn enable tighter constraints on halo mass and cosmological parameters. 
This is an interesting possibility and will be explored in the future.

In this paper, we assume the Gaussian statistics for simplicity. The results in Section~\ref{subsec:validate} 
imply that the non-Gaussian covariance contribution, 
which arises from 
nonlinear large-scale structure at lens redshift, 
 is not significant for 
length scales of $R>3\, h^{-1}\, M_{\odot}$ for ongoing imaging surveys. 
The non-Gaussian contribution might be more important for future surveys such as LSST which have a higher number density of source galaxies
and therefore have a more significant contribution of sample variance to the stacked lensing measurement.
We leave this question to our future work. For this, the method developed in \citet{TakadaHu:13} would be useful to extend the formulation to the stacked 
lensing profile. 

\section*{Acknowledgements}
We thank Bhuvnesh Jain,
Hironao Miyatake, Ryoma Murata, Takahiro Nishimichi, Masamune Oguri, and Ryuichi Takahashi
for useful discussions.
This work was supported by JSPS Grant-in-Aid for JSPS Research Fellow
Grant Number JP16J01512.
This work was supported in part by World Premier International
Research Center Initiative (WPI Initiative), MEXT, Japan, and JSPS
KAKENHI Grant Number JP26800093, JP23340061, JP26610058, JP15H03654,
JP15H05887, JP15H05892, JP15H05893, JP15K21733, and 17K14273.
Numerical simulations were carried out on Cray XC30
at the Center for Computational Astrophysics,
National Astronomical Observatory of Japan.

\appendix

\section{Derivation of the covariance matrix for the stacked lensing profile}
\label{app:cov}

In this section, we derive the covariance matrix for the power spectrum of the stacked lensing profile using the method 
in \cite{TakadaBridle:07}.
From Eq.~(\ref{eq:C_dsigma}), we express an estimator for the projected power spectrum in terms of 
the average of the Fourier modes as 
\begin{align}
\widehat{C}_{\dsigma}(k_i;z_l)\equiv& \frac{1}{N_{\rm mode}(k_i)\int_{z_{\rm cut}}^\infty\!\!\mathrm{d}z_s~p(z_s)w(z_l,z_s)}
\int_{z_{\rm cut}}^\infty\!\!\mathrm{d}z_s
~p(z_s) w(z_l,z_{s})\sum_{\bk; |\bk|\in k_i} 
\sigmacr(z_l,z_s) \tdelta^{\rm 2D}_{\rm h}(\bk)\tkappa(-\bk,z_s)\nonumber\\
=&\frac{1}{N_{\rm mode}(k_i)\avrg{w(z_l,z_s)}_{z_s}\int_{z_{\rm cut}}^\infty\!\!\mathrm{d}z_s~p(z_s)}
\int_{z_{\rm cut}}^\infty\!\!\mathrm{d}z_s
~p(z_s) w(z_l,z_{s})\sum_{\bk; |\bk|\in k_i} 
\sigmacr(z_l,z_s) \tdelta^{\rm 2D}_{\rm h}(\bk)\tkappa(-\bk,z_s)\nonumber\\
=&\frac{\bar{n}_{\rm tot}}{N_{\rm mode}(k_i)\avrg{w(z_l,z_s)}_{z_s}\bar{n}_{s}}
\int_{z_{\rm cut}}^\infty\!\!\mathrm{d}z_s
~p(z_s) w(z_l,z_{s})\sum_{\bk; |\bk|\in k_i} 
\sigmacr(z_l,z_s) \tdelta^{\rm 2D}_{\rm h}(\bk)\tkappa(-\bk,z_s)\nonumber\\
\simeq &\frac{1}{\bar{n}_{s}N_{\rm mode}(k_i)\avrg{w(z_l,z_s)}_{z_s}}\sum_{i_s;z_{i_s}>z_{\rm cut}}\bar{n}_{i_s}w(z_l,z_{i_s})\sum_{\bk; |\bk|\in k_i} 
\sigmacr(z_l,z_{i_s}) \tdelta^{\rm 2D}_{\rm h}(\bk)\tkappa(-\bk,z_{i_s}),
\end{align}	
where
we have used Eq.~(\ref{eq:ns_def}) to rewrite
$\bar{n}_{\rm tot}\int_{z_{\rm cut}}^\infty\!\!\mathrm{d}z_s~p(z_s)=\bar{n}_s$;
we have used that the redshift distribution of source galaxies can be approximated by a discrete summation: 
\begin{equation}
\int_{z_{\rm cut}}^\infty\!\!\mathrm{d}z_s~p(z_s)=\frac{1}{\bar{n}}_{\rm tot}\bar{n}_{\rm tot}
\int_{z_{\rm cut}}^\infty\!\!\mathrm{d}z_s~p(z_s)
\simeq \frac{1}{\bar{n}_{\rm tot}}\sum_{i_s; z_{i_s}>z_{\rm cut}} \bar{n}_{\rm tot}p(z_{i_s})\mathrm{d}z_s
=\frac{1}{\bar{n}_{\rm tot}}\sum_{i_s; z_{i_s}>z_{\rm cut}} \bar{n}_{i_s},
\end{equation}	
where $\bar{n}_{i_s}$ is the mean number density of source galaxies in the $i_s$-th redshift bin.
$N_{\rm mode}(k_i)$ is the number of Fourier modes used for the power spectrum estimation at the $k_i$-bin, defined as
\begin{equation}
N_{\rm mode}(k_i)\equiv \sum_{\bk; |\bk|\in k_i}\simeq \frac{2\pi k_i \Delta k_i}{(2\pi/\chi_l\Theta_S)^2}=\frac{\chi_l^2\Omega_Sk_i\Delta k_i}{2\pi}
=2\chi_l^2 f_{\rm sky} k_i\Delta k_i,
\end{equation}
where $\Omega_S$ is the survey area, and $f_{\rm sky}$ is the area fraction on the sky; $f_{\rm sky}\equiv \Omega_S/4\pi$.
Note that $\tkappa(\bk,z_s)$ and $\tdelta^{\rm 2D}_{\rm h}(\bk)$ are the ``observed'' fields including the contamination 
of shape noise and shot noise, respectively. 
The ensemble average of this estimator gives an unbiased estimate of the power spectrum of stacked lensing:
\begin{align}
\avrg{\hat{C}_\dsigma(k_i)}=&\frac{1}{\bar{n}_{s}N_{\rm mode}(k_i)\avrg{w(z_l,z_s)}_{z_s}}\sum_{i_s}\bar{n}_{i_s}w(z_l,z_s)\sum_{\bk; |\bk|\in k_i} 
\avrg{\sigmacr(z_l,z_{i_s}) \tdelta_{\rm h}(\bk)\tkappa(-\bk,z_{i_s})}\nonumber\\
=& \frac{1}{\bar{n}_{s}N_{\rm mode}(k_i)\avrg{w(z_l,z_s)}_{z_s}}\sum_{i_s}\bar{n}_{i_s}w(z_l,z_s)\sum_{\bk; |\bk|\in k_i} C_\dsigma(k)\nonumber\\
\simeq & \frac{1}{\bar{n}_{s}N_{\rm mode}(k_i)\avrg{w(z_l,z_s)}_{z_s}}C_\dsigma(k_i)
\sum_{i_s}\bar{n}_{i_s}w(z_l,z_s)\sum_{\bk; |\bk|\in k_i} = C_\dsigma(k_i).
\end{align}

The covariance matrix is defined as
\begin{align}
{\rm Cov}\left[\hat{C}_{\dsigma}(k_i),\hat{C}_\dsigma(k_j)\right]\equiv & \frac{1}{(\bar{n}_{s})^2N_{\rm mode}(k_i)N_{\rm mode}(k_j)(\avrg{w(z_l,z_s)}_{z_s})^2}
\sum_{i_s}\bar{n}_{i_s}w(z_l,z_{i_s})\sum_{j_s}\bar{n}_{j_s}w(z_l,z_{j_s})\nonumber\\
&\hspace{-2em}\times \sum_{\bk; |\bk|\in k_i}
\sum_{\bk'; |\bk'|\in k_j} \!\!\avrg{\sigmacr(z_l,z_{i_s})
\sigmacr(z_l,z_{j_s})\tdelta^{\rm 2D}_{\rm h}(\bk)\tkappa(-\bk,z_{i_s})
\tdelta^{\rm 2D}_{\rm h}(\bk')\tkappa(-\bk',z_{j_s})} - C_\dsigma(k_i)C_\dsigma(k_j).
\label{eq:cov_def}
\end{align}
Here, assuming that the cosmic shear field and the projected number density field of halos follow Gaussian statistics,  
the 4-point correlation function on the r.h.s. can be simplified as
\begin{align}
\avrg{\sigmacr(z_l,z_{i_s})\sigmacr(z_l,z_{j_s})\tdelta^{\rm 2D}_{\rm h}(\bk)\tkappa(-\bk,z_{i_s})
\tdelta^{\rm 2D}_{\rm h}(\bk')\tkappa(-\bk',z_{j_s})}=&
C_\dsigma(k; z_{i_s})C_\dsigma(k'; z_{j_s}) \nonumber\\
&\hspace{-22em}+ \avrg{\sigmacr(z_l,z_{j_s})\tdelta^{\rm 2D}_{\rm h}(\bk)
\tkappa(-\bk',z_{j_s})
}\avrg{\sigmacr(z_l,z_{i_s})\tdelta^{\rm 2D}_{\rm h}(\bk')
\tkappa(-\bk,z_{i_s})}
+ \avrg{\tdelta^{\rm 2D}_{\rm h}(\bk)\tdelta^{\rm 2D}_{\rm h}(\bk')}
\avrg{\sigmacr(z_l,z_{i_s})\sigmacr(z_l,z_{j_s})\tkappa(-\bk,z_{i_s})\tkappa(-\bk',z_{j_s})}\nonumber\\
&\hspace{-24em}= C_\dsigma(k)C_\dsigma(k') + C_\dsigma(k; z_{i_s})C_\dsigma(k; z_{j_s})\delta^K_{\bk-\bk'}
+C_{\rm hh}(k)\delta^K_{\bk+\bk'}
\avrg{\sigmacr(z_l,z_{i_s})\sigmacr(z_l,z_{j_s})\tkappa(-\bk,z_{i_s})\tkappa(-\bk',z_{j_s})},
\label{eq:cov_step1}
\end{align}
where $\delta^K_{\bk+\bk'}$ is the Kronecker-type delta function: $\delta^K_{\bk+\bk'}=1$ if $\bk+\bk'={\bf 0}$, and otherwise $\delta^K_{\bk+\bk'}=0$. 
Inserting the last term on the r.h.s. of the above equation into Eq.~(\ref{eq:cov_def}) leads to  
\begin{align}
&\hspace{0em}\frac{1}{(\bar{n}_{s})^2(\avrg{w(z_l,z_s)}_{z_s})^2}\sum_{i_s}\sum_{j_{s}}\bar{n}_{i_s}\bar{n}_{j_s}w(z_l,z_{i_s})w(z_l,z_{j_s})
\avrg{\sigmacr(z_l,z_{i_s})\sigmacr(z_l,z_{j_s})\tkappa^{\rm obs}(\bk;z_{i_s})\tkappa^{\rm obs}({\bk';z_{j_s}})}\nonumber\\
&\hspace{2em}=
\frac{1}{(\bar{n}_{s})^2(\avrg{w(z_l,z_s)}_{z_s})^2}\sum_{i_s}\sum_{j_{s}}\bar{n}_{i_s}\bar{n}_{j_s}w(z_l,z_{i_s})w(z_l,z_{j_s})
\sigmacr(z_l,z_{i_s})\sigmacr(z_l,z_{j_s})\left[
\avrg{\tkappa(\bk;z_{i_s})\tkappa({\bk';z_{j_s}})}
+\delta^K_{i_sj_s}\chi_l^2
\frac{\sigma_\epsilon^2}{\bar{n}_{i_s}}(2\pi)^2\delta^K_{\bk+\bk'}
\right]
\nonumber\\
&\hspace{2em}=
\frac{1}{(\avrg{w(z_l,z_s)}_{z_s})^2\left(\int_{z_{\rm cut}}^\infty\!\!\mathrm{d}z_s~p(z_s)\right)^2}
\int_{z_{\rm cut}}^\infty\!\!\mathrm{d}z_s~ p(z_s)w(z_l,z_s) 
\int_{z_{\rm cut}}^\infty\!\!\mathrm{d}z_s^\prime~ p(z_s^\prime)w(z_l,z_s^\prime)
\sigmacr(z_l,z_{s})\sigmacr(z_l,z_{s}^\prime)
\avrg{\tkappa(\bk;z_{s})\tkappa(\bk';z_{s}^\prime)}
\nonumber\\
&\hspace{12em}+
\frac{1}{\bar{n}_s(\avrg{w(z_l,z_s)}_{z_s})^2\int_{z_{\rm cut}}^\infty\!\mathrm{d}z_s~p(z_s)}
\int_{z_{\rm cut}}^\infty\!\!\mathrm{d}z_s~p(z_s)w(z_l,z_s)^2\sigmacr(z_l,z_s)^2
\chi_l^2\sigma_\epsilon^2(2\pi)^2\delta^K_{\bk+\bk'}
\label{eq:cov_step2}
\end{align}	
Using the Limber's approximation and the definition of the two-dimensional Fourier transform (e.g., Eq.~\ref{eq:2DFourier}), 
the first term of the above equation can be further simplified as
\begin{align}
&
\frac{1}{(\avrg{w(z_l,z_s)}_{z_s})^2\left(\int_{z_{\rm cut}}^\infty\!\!\mathrm{d}z_s~ p(z_s)\right)^2}
\int_{z_{\rm cut}}^\infty\!\!\mathrm{d}z_s~ p(z_s) w(z_l,z_s)
\int_{z_{\rm cut}}^\infty\!\!\mathrm{d}z_s^\prime~ p(z_s^\prime) w(z_l,z_s^\prime)
\sigmacr(z_l,z_{s})\sigmacr(z_l,z_{s}^\prime)
\avrg{\tkappa(\bk;z_{i_s})\tkappa({\bk';z_{j_s}})}
\nonumber\\
&\hspace{5em}\rightarrow
\frac{1}{(\avrg{w(z_l,z_s)}_{z_s})^2\left(\int_{z_{\rm cut}}^\infty\!\!\mathrm{d}z_s~ p(z_s)\right)^2}
\int_{z_{\rm cut}}^\infty\!\!\mathrm{d}z_s~ p(z_s) w(z_l,z_s)
\int_{z_{\rm cut}}^\infty\!\!\mathrm{d}z_s^\prime~ p(z_s^\prime)w(z_l,z_s^\prime)
\sigmacr(z_l,z_{s})\sigmacr(z_l,z_{s}^\prime)\nonumber\\
&\hspace{7em}\times
\int_{\chi_s}^\infty\!\!\mathrm{d}\chi~\int_{\chi_s^\prime}^\infty\!\!\mathrm{d}\chi'~\sigmacri(z,z_s)\sigmacri(z',z_s^\prime)
(\bar{\rho}_{\rm m0})^2\left(\frac{\chi_l}{\chi}\right)^2
\int\!\!\frac{\mathrm{d}k_\parallel}{2\pi}P_{\rm m}\left(k_s=\frac{\chi_l}{\chi}k;\chi,\chi'\right)
e^{ik_\parallel(\chi-\chi')}
\nonumber\\
&\hspace{5em}=
\frac{1}{(\avrg{w(z_l,z_s)}_{z_s})^2\left(\int_{z_{\rm cut}}^\infty\!\!\mathrm{d}z_s~ p(z_s)\right)^2}
\int_{z_{\rm cut}}^\infty\!\!\mathrm{d}z_s~ p(z_s) w(z_l,z_s)
\int_{z_{\rm cut}}^\infty\!\!\mathrm{d}z_s^\prime~ p(z_s^\prime)w(z_l,z_s^\prime)
\sigmacr(z_l,z_{s})\sigmacr(z_l,z_{s}^\prime)
\nonumber\\
&\hspace{8em}\times
\int_0^{\min\{\chi_s,\chi_s^\prime\}}\!\!\mathrm{d}\chi~\sigmacri(z,z_s)\sigmacri(z,z_s^\prime)
(\bar{\rho}_{\rm m0})^2
\left(\frac{\chi_l}{\chi}\right)^2
P_{\rm m}\left(k_s=\frac{\chi_l}{\chi}k;\chi\right)
\nonumber\\
&\hspace{5em}=
\frac{1}{(\avrg{w(z_l,z_s)}_{z_s})^2}
\int_0^\infty\!\!\mathrm{d}\chi~ \left[
\avrg{\sigmacr(z_l,z_s)\sigmacri(z,z_s)w(z_l,z_s)}_{z_s}\bar{\rho}_{\rm m0}\right]^2
\left(\frac{\chi_l}{\chi}\right)^2
P_{\rm m}\!\left(k_s=\frac{\chi_l}{\chi}k;\chi\right)
(2\pi)^2\delta_D^2(\bk+\bk')
\label{eq:sampcov_1st}
\end{align}
where 
\begin{align}
\avrg{\sigmacr(z_l,z_s)\sigmacri(z,z_s)w(z_l,z_s)}_{z_s}\equiv
\frac{1}{\int_{z_{\rm cut}}^\infty\!\!\mathrm{d}z_s~p(z_s)}\int_{z={\rm max}\{z(\chi),z_{\rm cut}\}}^\infty\!\!\mathrm{d}z_s~p(z_s)
w(z_l,z_s)
\sigmacr(z_l,z_s)\sigmacri(z,z_s)
\label{eq:sampcov_2nd}
\end{align}
The 2nd term of Eq.~(\ref{eq:cov_step2}) is simplified as
\begin{align}
\frac{1}{\bar{n}_{s}(\avrg{w(z_l,z_s)}_{z_s})^2\int_{z_{\rm cut}}^\infty\!\!\mathrm{d}z_s~ p(z_s)}
\int_{z_{\rm cut}}^\infty\!\!\mathrm{d}z_s~p(z_s)w(z_l,z_s)^2
\sigmacr(z_l,z_{s})^2
\chi_l^2\sigma_\epsilon^2&=&
\frac{\avrg{w(z_l,z_s)^2\sigmacr(z_l,z_s)^2}_{z_s}}{\bar{n}_{s}\left(\avrg{w(z_l,z_s)}_{z_s}\right)^2}
\chi_l^2\sigma_\epsilon^2
\end{align}
%
%
Therefore, inserting Eqs.~(\ref{eq:sampcov_1st}) and (\ref{eq:sampcov_2nd}) into Eq.~(\ref{eq:cov_def}), 
the covariance matrix for the stacked lensing power spectrum is expressed as
\begin{align}
{\rm Cov}\left[\hat{C}_\dsigma(k_i),\hat{C}_\dsigma(k_j)\right]=\frac{\delta^K_{ij}}{N_{\rm mode}(k_i)}
\left[C_\dsigma(k_i)^2+C^{\rm obs}_{\rm hh}(k_i)C^{\rm obs}_{\kappa\kappa,\sigmacr}\!(k_i)\right], \label{eq:cov_dSigma}
\end{align}
where 
\begin{eqnarray}
&&\hspace{-2em}N_{\rm mode}(k_i)=2 \chi_l^2 f_{\rm sky}k_i\Delta k_i\nonumber\\
&&\hspace{-2em}C^{\rm obs}_{\rm hh}=C_{\rm hh}(k)+\frac{\chi_l^2}{\bar{n}^{\rm 2D}_{\rm h}}\nonumber\\
&&\hspace{-2em}C^{\rm obs}_{\kappa\kappa,\sigmacr}(k_i)\equiv 
\frac{1}{(\avrg{w(z_l,z_s)}_{z_s})^2}
\int_0^\infty\!\!\mathrm{d}\chi~\avrg{
\sigmacr(z_l,z_s)\sigmacri(z,z_s)w(z_l,z_s)}_{z_s}^2
(\bar{\rho}_{\rm m0})^2
\left(\frac{\chi_l}{\chi}\right)^2
P_{\rm m}\!\left(k=\frac{\chi_l}{\chi}k_i; \chi\right)
+\frac{\avrg{w(z_l,z_s)^2\sigmacr(z_l,z_s)^2}_{z_s}}{\bar{n}_{s}\left(\avrg{w(z_l,z_s)}_{z_s}\right)^2}
\chi_l^2\sigma_\epsilon^2.\nonumber\\
\label{eq:c_kk_sigmacr}
\end{eqnarray}
From this equation, we can find that the covariance matrix in the shot noise dominated regime is given as
\beq
{\rm Cov}\left[\hat{C}_\dsigma(k_i),\hat{C}_\dsigma(k_j)\right]
\simeq \frac{\delta^K_{ij}}{N_{\rm mode}(k_i)}\frac{1}{\bar{n}_{\rm h}}
\frac{\avrg{w(z_l,z_s)^2\sigmacr(z_l,z_s)^2}}{\bar{n}_{s}\left(\avrg{w(z_l,z_s)}\right)^2}
\chi_l^2\sigma_\epsilon^2.
\eeq
On the other hand, in teh sample variance dominated regime, the covariance is approximated as
\beq
{\rm Cov}\left[\hat{C}_\dsigma(k_i),\hat{C}_\dsigma(k_j)\right]
\simeq \frac{\delta^K_{ij}}{N_{\rm mode}(k_i)}
\left[C_\dsigma(k_i)^2+C_{\rm hh}(k_i)\int_0^\infty\!\!\mathrm{d}\chi~\avrg{
\sigmacr(z_l,z_s)\sigmacri(z,z_s)\bar{\rho}_{\rm m0}}_{z_s,w}^2
\left(\frac{\chi_l}{\chi}\right)^2
P_{\rm m}\!\left(k_s=\frac{\chi_l}{\chi}k; \chi\right)\right]
\eeq


\section{Variance of excess surface mass density as a function of weight}
\label{app:mock_var_largescale}

\begin{figure*}
\centering
\includegraphics[width=0.45\columnwidth]{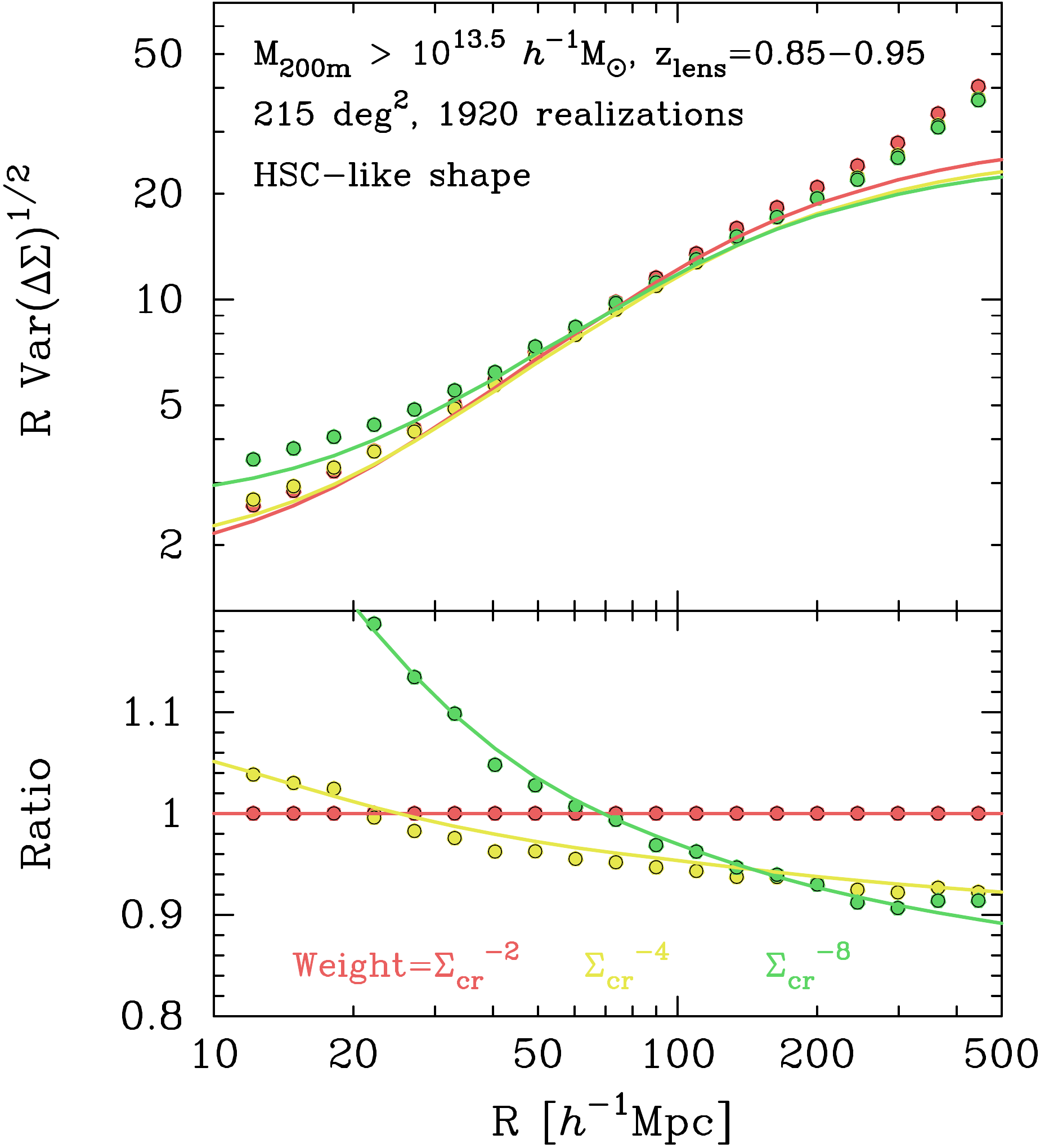}
\caption{
	Comparison of 
        mock results 
    and our analytic models for the  variance 
    of
    excess surface mass density estimator ($\Delta \Sigma$).
    The upper panel shows the variance
    of $\Delta \Sigma$ with the weight 
    of $\sigmacr^{-\alpha}$,
    while the bottom represents 
    the ratio normalized to the case 
    with the weight  $\sigmacr^{-2}$.
    Here we consider three cases of 
    $\alpha=2$, 4, and 8 
    and the HSC-type survey with 
    the survey coverage of 215 ${\rm deg}^2$.
    In this figure, we assume 
    a mass-limited foreground sample with
    mass of $M>10^{13.5}\,h^{-1} M_{\odot}$
    at redshift of $z_l = 0.85-0.95$.
    The different colored points show the mock results with different $\alpha$,
    while the lines are for our halo model predictions.
	}
\label{fig:var_dsigma_vary_weight}
\end{figure*} 

In this section, we use mock catalogs to 
examine how the signal-to-noise ratio of stacked excess survey mass density $\Delta \Sigma$ 
in the sample variance limited regime can be improved by using the weight  $\sigmacr^{-\alpha}$ $(\alpha > 2)$.
We utilize 1920 realizations of 
the HSC-type mock catalog in Section~\ref{subsec:validate}
with the sky coverage of 215 ${\rm deg}^{2}$.
For a foreground sample, we here consider
a mass-limited sample with 
$M_{\rm 200m}\ge10^{13.5}\, h^{-1}\, M_{\odot}$
at lens redshift $z_l=0.85-0.95$.
We then use the all source galaxies behind foreground objects, which is equivalent 
to $z_{\rm cut} = 0.95$.
When performing the mock stacked lensing anlysis, we correct for the observed $\Delta \Sigma$
by subtracting of the signal around random points.
As in section~\ref{subsec:validate},
we set the number of random points to be
10 times as large as that of foreground objects.
Since we are interested in the signal in the sample variance limited regime, we extend the outer radius in the mock analyses from 
$73.6\,h^{-1}\,{\rm Mpc}$ to $445.3\,h^{-1}\,{\rm Mpc}$ with the bin width of $\Delta\ln R=0.2$.

The upper panel in Figure~\ref{fig:var_dsigma_vary_weight}
shows the variances of $\Delta \Sigma$
with the weight of $\sigmacr^{-2}$, $\sigmacr^{-4}$ or $\sigmacr^{-8}$,
while the bottom represents the ratio normalized
to the case with the weight $\sigmacr^{-2}$.
In this figure, red, yellow, and green points
are the mock results with the weight $\sigmacr^{-2}$, $\sigmacr^{-4}$ and $\sigmacr^{-8}$, respectively.
The colored lines correspond to our halo model predictions (see Eqs.~\ref{eq:dsigma_real} and \ref{eq:cov_dsigma_real}).
We find that our model can provide a reasonable fit to the mock variances with the weight of
$\sigmacr^{-4}$ and $\sigmacr^{-8}$.
Around the boundary of survey geometry,
we expect the scaling of covariance with survey area breaks down due to 
the existence of super-survey modes 
\citep[see e.g.][]{Shirasakietal:17}.
Despite the sizable deviation at 
$R\simgt250\,h^{-1}\, {\rm Mpc}$
between the mock result and our model,
we confirm the reduction in the mock variance
as $\alpha$ increases 
over the wide range of radii.



\bibliographystyle{mnras}
\bibliography{refs} 

\begin{thebibliography}{}
\makeatletter
\relax
\def\mn@urlcharsother{\let\do\@makeother \do\$\do\&\do\#\do\^\do\_\do\%\do\~}
\def\mn@doi{\begingroup\mn@urlcharsother \@ifnextchar [ {\mn@doi@}
  {\mn@doi@[]}}
\def\mn@doi@[#1]#2{\def\@tempa{#1}\ifx\@tempa\@empty \href
  {http://dx.doi.org/#2} {doi:#2}\else \href {http://dx.doi.org/#2} {#1}\fi
  \endgroup}
\def\mn@eprint#1#2{\mn@eprint@#1:#2::\@nil}
\def\mn@eprint@arXiv#1{\href {http://arxiv.org/abs/#1} {{\tt arXiv:#1}}}
\def\mn@eprint@dblp#1{\href {http://dblp.uni-trier.de/rec/bibtex/#1.xml}
  {dblp:#1}}
\def\mn@eprint@#1:#2:#3:#4\@nil{\def\@tempa {#1}\def\@tempb {#2}\def\@tempc
  {#3}\ifx \@tempc \@empty \let \@tempc \@tempb \let \@tempb \@tempa \fi \ifx
  \@tempb \@empty \def\@tempb {arXiv}\fi \@ifundefined
  {mn@eprint@\@tempb}{\@tempb:\@tempc}{\expandafter \expandafter \csname
  mn@eprint@\@tempb\endcsname \expandafter{\@tempc}}}

\bibitem[\protect\citeauthoryear{{Assassi}, {Simonovi{\'c}}  \&
  {Zaldarriaga}}{{Assassi} et~al.}{2017}]{2017JCAP...11..054A}
{Assassi} V.,  {Simonovi{\'c}} M.,   {Zaldarriaga} M.,  2017, \mn@doi [\jcap]
  {10.1088/1475-7516/2017/11/054}, \href
  {http://adsabs.harvard.edu/abs/2017JCAP...11..054A} {11, 054}

\bibitem[\protect\citeauthoryear{{Baldauf}, {Smith}, {Seljak}  \&
  {Mandelbaum}}{{Baldauf} et~al.}{2010}]{2010PhRvD..81f3531B}
{Baldauf} T.,  {Smith} R.~E.,  {Seljak} U.,   {Mandelbaum} R.,  2010, \mn@doi
  [\prd] {10.1103/PhysRevD.81.063531}, \href
  {http://adsabs.harvard.edu/abs/2010PhRvD..81f3531B} {81, 063531}

\bibitem[\protect\citeauthoryear{{Becker}}{{Becker}}{2013}]{2013MNRAS.435..115B}
{Becker} M.~R.,  2013, \mn@doi [\mnras] {10.1093/mnras/stt1352}, \href
  {http://adsabs.harvard.edu/abs/2013MNRAS.435..115B} {435, 115}

\bibitem[\protect\citeauthoryear{{Behroozi}, {Wechsler}  \& {Wu}}{{Behroozi}
  et~al.}{2013}]{2013ApJ...762..109B}
{Behroozi} P.~S.,  {Wechsler} R.~H.,   {Wu} H.-Y.,  2013, \mn@doi [\apj]
  {10.1088/0004-637X/762/2/109}, \href
  {http://adsabs.harvard.edu/abs/2013ApJ...762..109B} {762, 109}

\bibitem[\protect\citeauthoryear{{Bernstein} \& {Jarvis}}{{Bernstein} \&
  {Jarvis}}{2002}]{BernsteinJarvis:02}
{Bernstein} G.~M.,  {Jarvis} M.,  2002, \mn@doi [\aj] {10.1086/338085}, \href
  {http://adsabs.harvard.edu/abs/2002AJ....123..583B} {123, 583}

\bibitem[\protect\citeauthoryear{{Bleem} et~al.,}{{Bleem}
  et~al.}{2015}]{2015ApJS..216...27B}
{Bleem} L.~E.,  et~al., 2015, \mn@doi [\apjs] {10.1088/0067-0049/216/2/27},
  \href {http://adsabs.harvard.edu/abs/2015ApJS..216...27B} {216, 27}

\bibitem[\protect\citeauthoryear{{Brainerd}, {Blandford}  \&
  {Smail}}{{Brainerd} et~al.}{1996}]{1996ApJ...466..623B}
{Brainerd} T.~G.,  {Blandford} R.~D.,   {Smail} I.,  1996, \mn@doi [\apj]
  {10.1086/177537}, \href {http://adsabs.harvard.edu/abs/1996ApJ...466..623B}
  {466, 623}

\bibitem[\protect\citeauthoryear{{Cacciato}, {van den Bosch}, {More}, {Li},
  {Mo}  \& {Yang}}{{Cacciato} et~al.}{2009}]{2009MNRAS.394..929C}
{Cacciato} M.,  {van den Bosch} F.~C.,  {More} S.,  {Li} R.,  {Mo} H.~J.,
  {Yang} X.,  2009, \mn@doi [\mnras] {10.1111/j.1365-2966.2008.14362.x}, \href
  {http://adsabs.harvard.edu/abs/2009MNRAS.394..929C} {394, 929}

\bibitem[\protect\citeauthoryear{{DES Collaboration} et~al.,}{{DES
  Collaboration} et~al.}{2017}]{2017arXiv170801530D}
{DES Collaboration} et~al., 2017, preprint, \href
  {http://adsabs.harvard.edu/abs/2017arXiv170801530D} {} (\mn@eprint {arXiv}
  {1708.01530})

\bibitem[\protect\citeauthoryear{{Diemer} \& {Kravtsov}}{{Diemer} \&
  {Kravtsov}}{2015}]{2015ApJ...799..108D}
{Diemer} B.,  {Kravtsov} A.~V.,  2015, \mn@doi [\apj]
  {10.1088/0004-637X/799/1/108}, \href
  {http://adsabs.harvard.edu/abs/2015ApJ...799..108D} {799, 108}

\bibitem[\protect\citeauthoryear{{Dodelson}}{{Dodelson}}{2003}]{DodelsonBook}
{Dodelson} S.,  2003, {Modern cosmology}

\bibitem[\protect\citeauthoryear{{Fischer} et~al.,}{{Fischer}
  et~al.}{2000}]{Fischeretal:00}
{Fischer} P.,  et~al., 2000, \mn@doi [\aj] {10.1086/301540}, \href
  {http://adsabs.harvard.edu/abs/2000AJ....120.1198F} {120, 1198}

\bibitem[\protect\citeauthoryear{{Gillis} et~al.,}{{Gillis}
  et~al.}{2013}]{2013MNRAS.431.1439G}
{Gillis} B.~R.,  et~al., 2013, \mn@doi [\mnras] {10.1093/mnras/stt274}, \href
  {http://adsabs.harvard.edu/abs/2013MNRAS.431.1439G} {431, 1439}

\bibitem[\protect\citeauthoryear{{G{\'o}rski}, {Hivon}, {Banday}, {Wandelt},
  {Hansen}, {Reinecke}  \& {Bartelmann}}{{G{\'o}rski}
  et~al.}{2005}]{2005ApJ...622..759G}
{G{\'o}rski} K.~M.,  {Hivon} E.,  {Banday} A.~J.,  {Wandelt} B.~D.,  {Hansen}
  F.~K.,  {Reinecke} M.,   {Bartelmann} M.,  2005, \mn@doi [\apj]
  {10.1086/427976}, \href {http://adsabs.harvard.edu/abs/2005ApJ...622..759G}
  {622, 759}

\bibitem[\protect\citeauthoryear{{Guzik} \& {Seljak}}{{Guzik} \&
  {Seljak}}{2002}]{2002MNRAS.335..311G}
{Guzik} J.,  {Seljak} U.,  2002, \mn@doi [\mnras]
  {10.1046/j.1365-8711.2002.05591.x}, \href
  {http://adsabs.harvard.edu/abs/2002MNRAS.335..311G} {335, 311}

\bibitem[\protect\citeauthoryear{{Hamana} \& {Mellier}}{{Hamana} \&
  {Mellier}}{2001}]{2001MNRAS.327..169H}
{Hamana} T.,  {Mellier} Y.,  2001, \mn@doi [\mnras]
  {10.1046/j.1365-8711.2001.04685.x}, \href
  {http://adsabs.harvard.edu/abs/2001MNRAS.327..169H} {327, 169}

\bibitem[\protect\citeauthoryear{{Hasselfield} et~al.,}{{Hasselfield}
  et~al.}{2013}]{2013JCAP...07..008H}
{Hasselfield} M.,  et~al., 2013, \mn@doi [\jcap]
  {10.1088/1475-7516/2013/07/008}, \href
  {http://adsabs.harvard.edu/abs/2013JCAP...07..008H} {7, 008}

\bibitem[\protect\citeauthoryear{{Hikage}, {Takada}  \& {Spergel}}{{Hikage}
  et~al.}{2012}]{Hikageetal:13}
{Hikage} C.,  {Takada} M.,   {Spergel} D.~N.,  2012, \mn@doi [\mnras]
  {10.1111/j.1365-2966.2011.19987.x}, \href
  {http://adsabs.harvard.edu/abs/2012MNRAS.419.3457H} {419, 3457}

\bibitem[\protect\citeauthoryear{{Hikage}, {Mandelbaum}, {Takada}  \&
  {Spergel}}{{Hikage} et~al.}{2013}]{2013MNRAS.435.2345H}
{Hikage} C.,  {Mandelbaum} R.,  {Takada} M.,   {Spergel} D.~N.,  2013, \mn@doi
  [\mnras] {10.1093/mnras/stt1446}, \href
  {http://adsabs.harvard.edu/abs/2013MNRAS.435.2345H} {435, 2345}

\bibitem[\protect\citeauthoryear{{Hinshaw} et~al.,}{{Hinshaw}
  et~al.}{2013}]{2013ApJS..208...19H}
{Hinshaw} G.,  et~al., 2013, \mn@doi [\apjs] {10.1088/0067-0049/208/2/19},
  \href {http://adsabs.harvard.edu/abs/2013ApJS..208...19H} {208, 19}

\bibitem[\protect\citeauthoryear{{Hoekstra}, {Yee}  \& {Gladders}}{{Hoekstra}
  et~al.}{2004}]{2004ApJ...606...67H}
{Hoekstra} H.,  {Yee} H.~K.~C.,   {Gladders} M.~D.,  2004, \mn@doi [\apj]
  {10.1086/382726}, \href {http://adsabs.harvard.edu/abs/2004ApJ...606...67H}
  {606, 67}

\bibitem[\protect\citeauthoryear{{Hudson}, {Gwyn}, {Dahle}  \&
  {Kaiser}}{{Hudson} et~al.}{1998}]{1998ApJ...503..531H}
{Hudson} M.~J.,  {Gwyn} S.~D.~J.,  {Dahle} H.,   {Kaiser} N.,  1998, \mn@doi
  [\apj] {10.1086/306026}, \href
  {http://adsabs.harvard.edu/abs/1998ApJ...503..531H} {503, 531}

\bibitem[\protect\citeauthoryear{{Jeong}, {Komatsu}  \& {Jain}}{{Jeong}
  et~al.}{2009}]{2009PhRvD..80l3527J}
{Jeong} D.,  {Komatsu} E.,   {Jain} B.,  2009, \mn@doi [\prd]
  {10.1103/PhysRevD.80.123527}, \href
  {http://adsabs.harvard.edu/abs/2009PhRvD..80l3527J} {80, 123527}

\bibitem[\protect\citeauthoryear{{Johnston} et~al.,}{{Johnston}
  et~al.}{2007}]{2007arXiv0709.1159J}
{Johnston} D.~E.,  et~al., 2007, preprint, \href
  {http://adsabs.harvard.edu/abs/2007arXiv0709.1159J} {} (\mn@eprint {arXiv}
  {0709.1159})

\bibitem[\protect\citeauthoryear{{Joudaki} et~al.,}{{Joudaki}
  et~al.}{2018}]{2018MNRAS.474.4894J}
{Joudaki} S.,  et~al., 2018, \mn@doi [\mnras] {10.1093/mnras/stx2820}, \href
  {http://adsabs.harvard.edu/abs/2018MNRAS.474.4894J} {474, 4894}

\bibitem[\protect\citeauthoryear{{Krause} \& {Eifler}}{{Krause} \&
  {Eifler}}{2017}]{KrauseEifler:17}
{Krause} E.,  {Eifler} T.,  2017, \mn@doi [\mnras] {10.1093/mnras/stx1261},
  \href {http://adsabs.harvard.edu/abs/2017MNRAS.470.2100K} {470, 2100}

\bibitem[\protect\citeauthoryear{{Kwan} et~al.,}{{Kwan}
  et~al.}{2017}]{2017MNRAS.464.4045K}
{Kwan} J.,  et~al., 2017, \mn@doi [\mnras] {10.1093/mnras/stw2464}, \href
  {http://adsabs.harvard.edu/abs/2017MNRAS.464.4045K} {464, 4045}

\bibitem[\protect\citeauthoryear{{Limber}}{{Limber}}{1954}]{Limber:54}
{Limber} D.~N.,  1954, \mn@doi [\apj] {10.1086/145870}, \href
  {http://adsabs.harvard.edu/abs/1954ApJ...119..655L} {119, 655}

\bibitem[\protect\citeauthoryear{{Mandelbaum} et~al.,}{{Mandelbaum}
  et~al.}{2005a}]{2005MNRAS.361.1287M}
{Mandelbaum} R.,  et~al., 2005a, \mn@doi [\mnras]
  {10.1111/j.1365-2966.2005.09282.x}, \href
  {http://adsabs.harvard.edu/abs/2005MNRAS.361.1287M} {361, 1287}

\bibitem[\protect\citeauthoryear{{Mandelbaum}, {Tasitsiomi}, {Seljak},
  {Kravtsov}  \& {Wechsler}}{{Mandelbaum} et~al.}{2005b}]{Mandelbaumetal:05}
{Mandelbaum} R.,  {Tasitsiomi} A.,  {Seljak} U.,  {Kravtsov} A.~V.,
  {Wechsler} R.~H.,  2005b, \mn@doi [\mnras]
  {10.1111/j.1365-2966.2005.09417.x}, \href
  {http://adsabs.harvard.edu/abs/2005MNRAS.362.1451M} {362, 1451}

\bibitem[\protect\citeauthoryear{{Mandelbaum}, {Seljak}, {Kauffmann}, {Hirata}
  \& {Brinkmann}}{{Mandelbaum} et~al.}{2006a}]{2006MNRAS.368..715M}
{Mandelbaum} R.,  {Seljak} U.,  {Kauffmann} G.,  {Hirata} C.~M.,   {Brinkmann}
  J.,  2006a, \mn@doi [\mnras] {10.1111/j.1365-2966.2006.10156.x}, \href
  {http://adsabs.harvard.edu/abs/2006MNRAS.368..715M} {368, 715}

\bibitem[\protect\citeauthoryear{{Mandelbaum}, {Seljak}, {Cool}, {Blanton},
  {Hirata}  \& {Brinkmann}}{{Mandelbaum} et~al.}{2006b}]{Mandelbaumetal:06}
{Mandelbaum} R.,  {Seljak} U.,  {Cool} R.~J.,  {Blanton} M.,  {Hirata} C.~M.,
  {Brinkmann} J.,  2006b, \mn@doi [\mnras] {10.1111/j.1365-2966.2006.10906.x},
  \href {http://adsabs.harvard.edu/abs/2006MNRAS.372..758M} {372, 758}

\bibitem[\protect\citeauthoryear{{Mandelbaum}, {Seljak}, {Baldauf}  \&
  {Smith}}{{Mandelbaum} et~al.}{2010}]{2010MNRAS.405.2078M}
{Mandelbaum} R.,  {Seljak} U.,  {Baldauf} T.,   {Smith} R.~E.,  2010, \mn@doi
  [\mnras] {10.1111/j.1365-2966.2010.16619.x}, \href
  {http://adsabs.harvard.edu/abs/2010MNRAS.405.2078M} {405, 2078}

\bibitem[\protect\citeauthoryear{{Mandelbaum}, {Slosar}, {Baldauf}, {Seljak},
  {Hirata}, {Nakajima}, {Reyes}  \& {Smith}}{{Mandelbaum}
  et~al.}{2013}]{Mandelbaumetal:13}
{Mandelbaum} R.,  {Slosar} A.,  {Baldauf} T.,  {Seljak} U.,  {Hirata} C.~M.,
  {Nakajima} R.,  {Reyes} R.,   {Smith} R.~E.,  2013, \mn@doi [\mnras]
  {10.1093/mnras/stt572}, \href
  {http://adsabs.harvard.edu/abs/2013MNRAS.432.1544M} {432, 1544}

\bibitem[\protect\citeauthoryear{{McKay} et~al.,}{{McKay}
  et~al.}{2001}]{2001astro.ph..8013M}
{McKay} T.~A.,  et~al., 2001, ArXiv Astrophysics e-prints, \href
  {http://adsabs.harvard.edu/abs/2001astro.ph..8013M} {}

\bibitem[\protect\citeauthoryear{{Miyatake} et~al.,}{{Miyatake}
  et~al.}{2015}]{Miyatakeetal:15}
{Miyatake} H.,  et~al., 2015, \mn@doi [\apj] {10.1088/0004-637X/806/1/1}, \href
  {http://adsabs.harvard.edu/abs/2015ApJ...806....1M} {806, 1}

\bibitem[\protect\citeauthoryear{{More}, {Miyatake}, {Mandelbaum}, {Takada},
  {Spergel}, {Brownstein}  \& {Schneider}}{{More}
  et~al.}{2015}]{2015ApJ...806....2M}
{More} S.,  {Miyatake} H.,  {Mandelbaum} R.,  {Takada} M.,  {Spergel} D.~N.,
  {Brownstein} J.~R.,   {Schneider} D.~P.,  2015, \mn@doi [\apj]
  {10.1088/0004-637X/806/1/2}, \href
  {http://adsabs.harvard.edu/abs/2015ApJ...806....2M} {806, 2}

\bibitem[\protect\citeauthoryear{{Murata}, {Nishimichi}, {Takada}, {Miyatake},
  {Shirasaki}, {More}, {Takahashi}  \& {Osato}}{{Murata}
  et~al.}{2017}]{Murataetal:17}
{Murata} R.,  {Nishimichi} T.,  {Takada} M.,  {Miyatake} H.,  {Shirasaki} M.,
  {More} S.,  {Takahashi} R.,   {Osato} K.,  2017, preprint, \href
  {http://adsabs.harvard.edu/abs/2017arXiv170701907M} {} (\mn@eprint {arXiv}
  {1707.01907})

\bibitem[\protect\citeauthoryear{{Navarro}, {Frenk}  \& {White}}{{Navarro}
  et~al.}{1997}]{1997ApJ...490..493N}
{Navarro} J.~F.,  {Frenk} C.~S.,   {White} S.~D.~M.,  1997, \mn@doi [\apj]
  {10.1086/304888}, \href {http://adsabs.harvard.edu/abs/1997ApJ...490..493N}
  {490, 493}

\bibitem[\protect\citeauthoryear{{Oguri} \& {Takada}}{{Oguri} \&
  {Takada}}{2011}]{OguriTakada:11}
{Oguri} M.,  {Takada} M.,  2011, \mn@doi [\prd] {10.1103/PhysRevD.83.023008},
  \href {http://adsabs.harvard.edu/abs/2011PhRvD..83b3008O} {83, 023008}

\bibitem[\protect\citeauthoryear{{Oguri} et~al.,}{{Oguri}
  et~al.}{2018}]{2018PASJ...70S..20O}
{Oguri} M.,  et~al., 2018, \mn@doi [\pasj] {10.1093/pasj/psx042}, \href
  {http://adsabs.harvard.edu/abs/2018PASJ...70S..20O} {70, S20}

\bibitem[\protect\citeauthoryear{{Okabe}, {Takada}, {Umetsu}, {Futamase}  \&
  {Smith}}{{Okabe} et~al.}{2010}]{Okabeetal:10}
{Okabe} N.,  {Takada} M.,  {Umetsu} K.,  {Futamase} T.,   {Smith} G.~P.,  2010,
  Publ.~Soc.~Astron.~Japan, \href
  {http://adsabs.harvard.edu/abs/2010PASJ...62..811O} {62, 811}

\bibitem[\protect\citeauthoryear{{Okabe}, {Smith}, {Umetsu}, {Takada}  \&
  {Futamase}}{{Okabe} et~al.}{2013}]{2013ApJ...769L..35O}
{Okabe} N.,  {Smith} G.~P.,  {Umetsu} K.,  {Takada} M.,   {Futamase} T.,  2013,
  \mn@doi [\apjl] {10.1088/2041-8205/769/2/L35}, \href
  {http://adsabs.harvard.edu/abs/2013ApJ...769L..35O} {769, L35}

\bibitem[\protect\citeauthoryear{{Prat} et~al.,}{{Prat}
  et~al.}{2017}]{PratetalDESgglensing:17}
{Prat} J.,  et~al., 2017, preprint, \href
  {http://adsabs.harvard.edu/abs/2017arXiv170801537P} {} (\mn@eprint {arXiv}
  {1708.01537})

\bibitem[\protect\citeauthoryear{{Reyes}, {Mandelbaum}, {Seljak}, {Baldauf},
  {Gunn}, {Lombriser}  \& {Smith}}{{Reyes} et~al.}{2010}]{Reyesetal:10}
{Reyes} R.,  {Mandelbaum} R.,  {Seljak} U.,  {Baldauf} T.,  {Gunn} J.~E.,
  {Lombriser} L.,   {Smith} R.~E.,  2010, \mn@doi [\nat] {10.1038/nature08857},
  \href {http://adsabs.harvard.edu/abs/2010Natur.464..256R} {464, 256}

\bibitem[\protect\citeauthoryear{{Schaan}, {Krause}, {Eifler}, {Dor{\'e}},
  {Miyatake}, {Rhodes}  \& {Spergel}}{{Schaan} et~al.}{2016}]{Schaanetal:16}
{Schaan} E.,  {Krause} E.,  {Eifler} T.,  {Dor{\'e}} O.,  {Miyatake} H.,
  {Rhodes} J.,   {Spergel} D.~N.,  2016, preprint, \href
  {http://adsabs.harvard.edu/abs/2016arXiv160701761S} {} (\mn@eprint {arXiv}
  {1607.01761})

\bibitem[\protect\citeauthoryear{{Seljak} et~al.,}{{Seljak}
  et~al.}{2005}]{Seljaketal:05}
{Seljak} U.,  et~al., 2005, \mn@doi [\prd] {10.1103/PhysRevD.71.043511}, \href
  {http://adsabs.harvard.edu/abs/2005PhRvD..71d3511S} {71, 043511}

\bibitem[\protect\citeauthoryear{{Sheldon} et~al.,}{{Sheldon}
  et~al.}{2004}]{2004AJ....127.2544S}
{Sheldon} E.~S.,  et~al., 2004, \mn@doi [\aj] {10.1086/383293}, \href
  {http://adsabs.harvard.edu/abs/2004AJ....127.2544S} {127, 2544}

\bibitem[\protect\citeauthoryear{{Shirasaki}, {Hamana}  \&
  {Yoshida}}{{Shirasaki} et~al.}{2015}]{2015MNRAS.453.3043S}
{Shirasaki} M.,  {Hamana} T.,   {Yoshida} N.,  2015, \mn@doi [\mnras]
  {10.1093/mnras/stv1854}, \href
  {http://adsabs.harvard.edu/abs/2015MNRAS.453.3043S} {453, 3043}

\bibitem[\protect\citeauthoryear{Shirasaki, Takada, Miyatake, Takahashi,
  Hamana, Nishimichi  \& Murata}{Shirasaki et~al.}{2017}]{Shirasakietal:17}
Shirasaki M.,  Takada M.,  Miyatake H.,  Takahashi R.,  Hamana T.,  Nishimichi
  T.,   Murata R.,  2017, \mn@doi [Mon. Not. Roy. Astron. Soc.]
  {10.1093/mnras/stx1477}, 470, 3476

\bibitem[\protect\citeauthoryear{{Singh}, {Mandelbaum}, {Seljak}, {Slosar}  \&
  {Vazquez Gonzalez}}{{Singh} et~al.}{2016}]{2016arXiv161100752S}
{Singh} S.,  {Mandelbaum} R.,  {Seljak} U.,  {Slosar} A.,   {Vazquez Gonzalez}
  J.,  2016, preprint, \href
  {http://adsabs.harvard.edu/abs/2016arXiv161100752S} {} (\mn@eprint {arXiv}
  {1611.00752})

\bibitem[\protect\citeauthoryear{{Takada} \& {Bridle}}{{Takada} \&
  {Bridle}}{2007}]{TakadaBridle:07}
{Takada} M.,  {Bridle} S.,  2007, \mn@doi [New Journal of Physics]
  {10.1088/1367-2630/9/12/446}, \href
  {http://adsabs.harvard.edu/abs/2007NJPh....9..446T} {9, 446}

\bibitem[\protect\citeauthoryear{{Takada} \& {Hu}}{{Takada} \&
  {Hu}}{2013}]{TakadaHu:13}
{Takada} M.,  {Hu} W.,  2013, \mn@doi [\prd] {10.1103/PhysRevD.87.123504},
  \href {http://adsabs.harvard.edu/abs/2013PhRvD..87l3504T} {87, 123504}

\bibitem[\protect\citeauthoryear{{Takada} \& {Jain}}{{Takada} \&
  {Jain}}{2004}]{TakadaJain:04}
{Takada} M.,  {Jain} B.,  2004, \mn@doi [\mnras]
  {10.1111/j.1365-2966.2004.07410.x}, \href
  {http://adsabs.harvard.edu/abs/2004MNRAS.348..897T} {348, 897}

\bibitem[\protect\citeauthoryear{{Takada} \& {Jain}}{{Takada} \&
  {Jain}}{2009}]{TakadaJain:09}
{Takada} M.,  {Jain} B.,  2009, \mn@doi [\mnras]
  {10.1111/j.1365-2966.2009.14504.x}, \href
  {http://adsabs.harvard.edu/abs/2009MNRAS.395.2065T} {395, 2065}

\bibitem[\protect\citeauthoryear{{Takada} et~al.,}{{Takada}
  et~al.}{2014}]{Takadaetal:14}
{Takada} M.,  et~al., 2014, \mn@doi [\pasj] {10.1093/pasj/pst019}, \href
  {http://adsabs.harvard.edu/abs/2014PASJ...66R...1T} {66, R1}

\bibitem[\protect\citeauthoryear{{Takahashi}, {Sato}, {Nishimichi}, {Taruya}
  \& {Oguri}}{{Takahashi} et~al.}{2012}]{2012ApJ...761..152T}
{Takahashi} R.,  {Sato} M.,  {Nishimichi} T.,  {Taruya} A.,   {Oguri} M.,
  2012, \mn@doi [\apj] {10.1088/0004-637X/761/2/152}, \href
  {http://adsabs.harvard.edu/abs/2012ApJ...761..152T} {761, 152}

\bibitem[\protect\citeauthoryear{{Takahashi}, {Hamana}, {Shirasaki},
  {Namikawa}, {Nishimichi}, {Osato}  \& {Shiroyama}}{{Takahashi}
  et~al.}{2017}]{2017ApJ...850...24T}
{Takahashi} R.,  {Hamana} T.,  {Shirasaki} M.,  {Namikawa} T.,  {Nishimichi}
  T.,  {Osato} K.,   {Shiroyama} K.,  2017, \mn@doi [\apj]
  {10.3847/1538-4357/aa943d}, \href
  {http://adsabs.harvard.edu/abs/2017ApJ...850...24T} {850, 24}

\bibitem[\protect\citeauthoryear{{Tinker}, {Kravtsov}, {Klypin}, {Abazajian},
  {Warren}, {Yepes}, {Gottl{\"o}ber}  \& {Holz}}{{Tinker}
  et~al.}{2008}]{2008ApJ...688..709T}
{Tinker} J.,  {Kravtsov} A.~V.,  {Klypin} A.,  {Abazajian} K.,  {Warren} M.,
  {Yepes} G.,  {Gottl{\"o}ber} S.,   {Holz} D.~E.,  2008, \mn@doi [\apj]
  {10.1086/591439}, \href {http://adsabs.harvard.edu/abs/2008ApJ...688..709T}
  {688, 709}

\bibitem[\protect\citeauthoryear{{Tinker}, {Robertson}, {Kravtsov}, {Klypin},
  {Warren}, {Yepes}  \& {Gottl{\"o}ber}}{{Tinker}
  et~al.}{2010}]{2010ApJ...724..878T}
{Tinker} J.~L.,  {Robertson} B.~E.,  {Kravtsov} A.~V.,  {Klypin} A.,  {Warren}
  M.~S.,  {Yepes} G.,   {Gottl{\"o}ber} S.,  2010, \mn@doi [\apj]
  {10.1088/0004-637X/724/2/878}, \href
  {http://adsabs.harvard.edu/abs/2010ApJ...724..878T} {724, 878}

\bibitem[\protect\citeauthoryear{{Velander} et~al.,}{{Velander}
  et~al.}{2014}]{2014MNRAS.437.2111V}
{Velander} M.,  et~al., 2014, \mn@doi [\mnras] {10.1093/mnras/stt2013}, \href
  {http://adsabs.harvard.edu/abs/2014MNRAS.437.2111V} {437, 2111}

\bibitem[\protect\citeauthoryear{{de Putter} \& {Takada}}{{de Putter} \&
  {Takada}}{2010}]{dePutterTakada:10}
{de Putter} R.,  {Takada} M.,  2010, \mn@doi [\prd]
  {10.1103/PhysRevD.82.103522}, \href
  {http://adsabs.harvard.edu/abs/2010PhRvD..82j3522D} {82, 103522}

\bibitem[\protect\citeauthoryear{{van Uitert} et~al.,}{{van Uitert}
  et~al.}{2017}]{2017arXiv170605004V}
{van Uitert} E.,  et~al., 2017, preprint, \href
  {http://adsabs.harvard.edu/abs/2017arXiv170605004V} {} (\mn@eprint {arXiv}
  {1706.05004})

\makeatother
\end{thebibliography}





\bsp	
\label{lastpage}
\end{document}